\documentclass[screen,sigconf]{acmart}

\usepackage{enumitem}

\copyrightyear{2023}
\acmYear{2023}
\setcopyright{rightsretained}
\acmConference[WWW '23]{Proceedings of the ACM Web Conference 2023}{April 30-May 4, 2023}{Austin, TX, USA}
\acmBooktitle{Proceedings of the ACM Web Conference 2023 (WWW '23), April 30-May 4, 2023, Austin, TX, USA}
\acmDOI{10.1145/3543507.3583503}
\acmISBN{978-1-4503-9416-1/23/04}

\usepackage{amsmath}
\usepackage{multirow}
\usepackage{afterpage}
\usepackage{xspace}
\newcommand{\model}{\textsf{CODER}\xspace}
\newcommand{\modelbf}{\textbf{CODER}\xspace}

\begin{document}

\title{Code Recommendation for Open Source Software Developers}

\author{Yiqiao Jin}
\orcid{0000-0002-6974-5970}
\affiliation{
  \institution{Georgia Institute of Technology}
  \streetaddress{North Ave NW}
  \city{Atlanta}
  \state{GA}
  \country{USA}
  \postcode{30332}
}
\email{yjin328@gatech.edu}

\author{Yunsheng Bai}
\orcid{0000-0003-1623-6184}
\affiliation{
 \institution{University of California, Los Angeles}
 \streetaddress{405 Hilgard Avenue}
 \city{Los Angeles}
 \state{CA}
 \country{USA}}
\email{yba@cs.ucla.edu}

\author{Yanqiao Zhu}
\orcid{0000-0003-2205-5304}
\affiliation{
 \institution{University of California, Los Angeles}
 \streetaddress{405 Hilgard Avenue}
 \city{Los Angeles}
 \state{CA}
 \country{USA}}
\email{yzhu@cs.ucla.edu}

\author{Yizhou Sun}
\orcid{0000-0003-1812-6843}
\affiliation{
 \institution{University of California, Los Angeles}
 \streetaddress{405 Hilgard Avenue}
 \city{Los Angeles}
 \state{CA}
 \country{USA}}
\email{yzsun@cs.ucla.edu}

\author{Wei Wang}
\orcid{0000-0002-8180-2886}
\affiliation{
    \institution{University of California, Los Angeles}
    \streetaddress{405 Hilgard Avenue}
    \city{Los Angeles}
    \state{CA}
    \country{USA}}
\email{weiwang@cs.ucla.edu}

\renewcommand{\shortauthors}{Jin et al.}

\newcommand{\yba}[1]{\textcolor{blue}{#1}}
\newcommand{\yiq}[1]{\textcolor{orange}{#1}}
\newcommand{\todo}[1]{\textcolor{green}{#1}}
\newcommand{\YS}[1]{{\bf\color{red}[{\sc YS:} #1]}}
\newcommand{\ww}[1]{{\bf\color{red}[{\sc ww:} #1]}}

\begin{abstract}
Open Source Software (OSS) is forming the spines of technology infrastructures, attracting millions of talents to contribute. Notably, it is challenging and critical to consider both the developers' interests and the semantic features of the project code to recommend appropriate development tasks to OSS developers. In this paper, we formulate the novel problem of code recommendation, whose purpose is to predict the future contribution behaviors of developers given their interaction history, the semantic features of source code, and the hierarchical file structures of projects. We introduce \model, a novel graph-based \underline{CODE} \underline{R}ecommendation framework for open source software developers, which accounts for the complex interactions among multiple parties within the system. \model jointly models microscopic user-code interactions and macroscopic user-project interactions via a heterogeneous graph and further bridges the two levels of information through aggregation on file-structure graphs that reflect the project hierarchy. Moreover, to overcome the lack of reliable benchmarks, we construct three large-scale datasets to facilitate future research in this direction. Extensive experiments show that our \model framework achieves superior performance under various experimental settings, including intra-project, cross-project, and cold-start recommendation. 
\end{abstract}



\maketitle

\section{Introduction}


Open Source Software (OSS) is becoming increasingly popular in software engineering~\cite{venkataramani2013discovery, jiang2017open}.
As contribution to OSS projects is highly democratized~\cite{yu2016reviewer}, these projects attract millions of developers with diverse expertise and efficiently crowd-source the project development to a larger community of developers beyond the project's major personnel ~\cite{mcdonald2013performance, jiang2017open}. 
For instance, GitHub, one of the most successful platforms for developing and hosting OSS projects, has over 83 million users and 200 million repositories~\cite{githubarchive}.

\begin{figure}
  \centering
  \includegraphics[width=\linewidth]{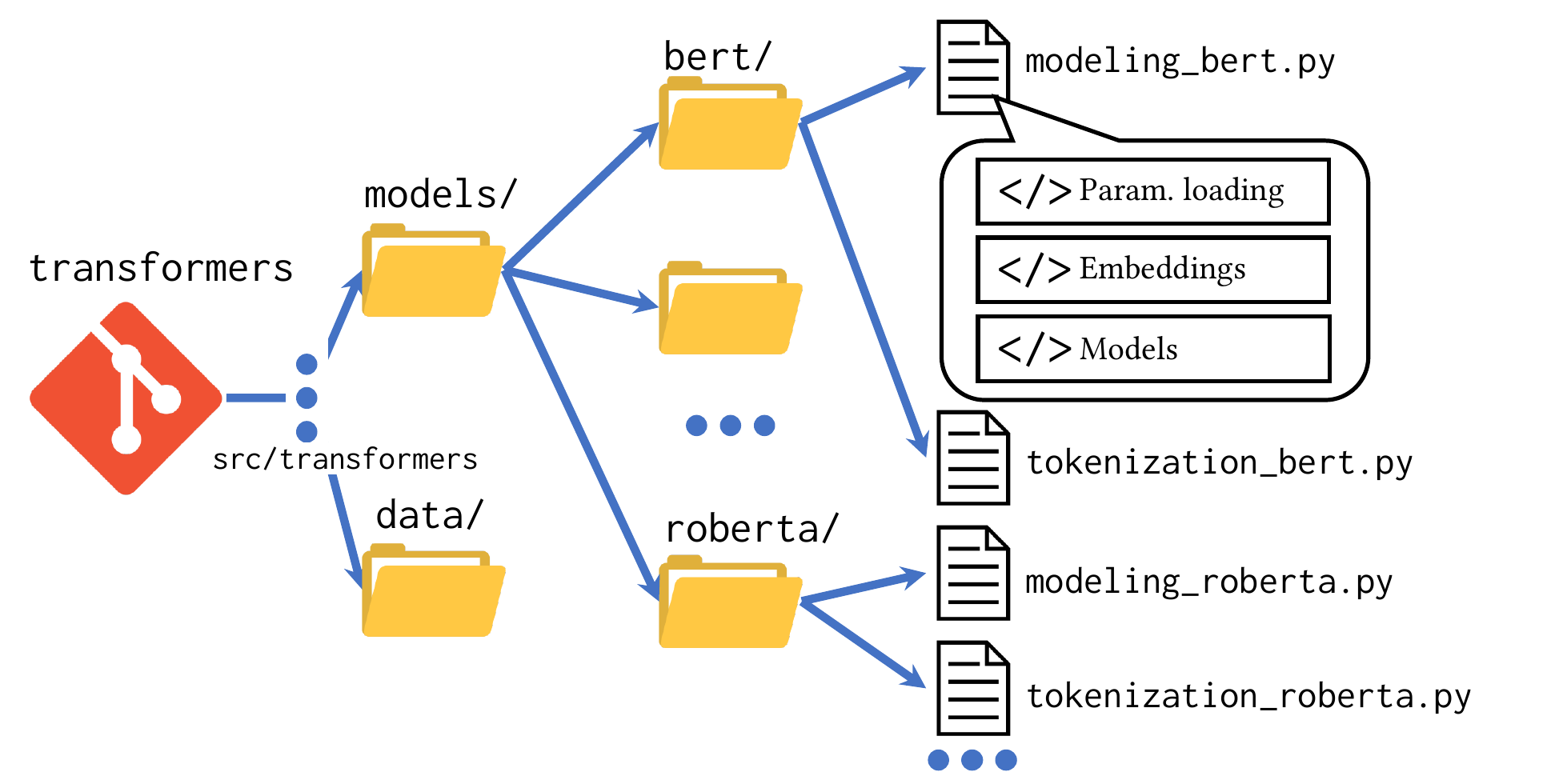}
  \caption{
    An example of the \texttt{transformers} repository. OSS projects under similar topics usually adopt similar naming conventions and file structures, which can be seen as knowledge transferable across projects.
  }
  \label{fig:motivating_Example}
\end{figure}

Community support and teamwork are major driving forces behind open source projects~\cite{mcdonald2013performance}. OSS projects are usually developed in a collaborative manner~\cite{borges2016understanding}, whereas collaboration in OSS is especially challenging. OSS projects are of large scales and usually contain numerous project files written in diverse programming languages~\cite{coelho2020github}.
According to statistics, the most popular 500 GitHub projects contain an average of 2,582 project files, 573 directories, and 360 contributors. Meanwhile, there are more than 300 programming languages on GitHub, 67 of which are actively being used~\cite{githublang2022, octoverse2016}.
For \textbf{project maintainers}, it is both difficult and time-consuming to find competent contributors within a potentially large candidate pool. 
For \textbf{OSS developers}, recommending personalized development tasks according to their project experience and expertise can significantly boost their motivation and reduce their cognitive loads of manually checking the project files.
As contribution in OSS is voluntary, developers that fail to find meaningful tasks are likely to quit the project development~\cite{steinmacher2014preliminary}. 
Therefore, an efficient system for automatically matching source code 
with potential contributors is being called for by both the project core team and the potential contributors to reduce their burden. 

To solve the above issues, in this paper, we for the first time introduce the novel problem of code recommendation for OSS developers. As shown in Fig.~\ref{fig:model}, this task recommends code in the form of project files to potentially suitable contributors. 
It is noteworthy that code recommendation has several unique challenges such that traditional recommender models are not directly applicable.

Firstly, OSS projects contain multimodal interactions among users, projects, and code files. For example, OSS development contains user-code interactions, such as commits that depict microscopic behaviors of users, and user-project interactions, such as forks and stars that exhibit users' macroscopic preferences and interests on projects.
Also, the contribution relationships are often extremely sparse, due to the significant efforts required to make a single contribution to OSS projects. 
Therefore, directly modeling the contribution behavior as in traditional collaborative filtering approaches will inevitably lead to inaccurate user/item representations and suboptimal performances.

Secondly, in the software engineering domain, code files in a project are often organized in a hierarchical structure~\cite{yu2021graph}.
Fig.~\ref{fig:motivating_Example} shows an example of the famous \path{huggingface/transformers} repository~\cite{wolf2020transformers}. The \path{src} directory usually contains the major source code for a project. The \path{data} and \path{models} subdirectories usually include functions for data generation and model implementations, respectively.
Such a structural organization of the OSS project reveals semantic relations among code snippets, which are helpful for developers to transfer existing code from other projects to their development. 
Traditional methods usually ignore such item-wise hierarchical relationships and, as a result, are incapable of connecting rich semantic features in code files with their project-level structures, which is required 
for accurate code recommendation. 

Thirdly, most existing benchmarks involving recommendation for softwares only consider limited user-item behaviors~\cite{di2017software, hu2018summarizing}, are of small scales~\cite{nguyen2019focus, nguyen2016api}, or contain only certain languages such as Python~\cite{he2021pyart, wan2018improving, miceli2017parallel} or Java~\cite{nguyen2019focus, di2017software, hu2018summarizing}, which renders the evaluation of different recommendation models difficult or not realistic. 

To overcome the above challenges, we propose \modelbf, a \underline{CODE} \underline{R}ecommendation framework for open source software developers that matches project files with potential contributors. 
As shown in Fig.~\ref{fig:model}, \model treats users, code files, and projects as nodes and jointly models the microscopic user-code interactions and macroscopic user-project interactions in a heterogeneous graph.
Furthermore, \model bridges these two levels of information through message aggregation on the file structure graphs that reflect the hierarchical relationships among graph nodes.
Additionally, since there is a lack of benchmark datasets for the code recommendation task, we build three large-scale datasets from open software development websites. These datasets cover diverse subtopics in computer science and contain up to 2 million fine-grained user-file interactions.
Overall, our contributions are summarized as follows:
\begin{itemize}[wide]
    \item We for the first time introduce the problem of code recommendation, whose purpose is to recommend appropriate development tasks to developers, given the interaction history of developers, the semantic features of source code, and hierarchical structures of projects.
    \item We propose \model, an end-to-end framework that jointly models structural and semantic features of source code as well as multiple types of user behaviors for improving the matching task.
    \item We construct three large-scale multi-modal datasets for code recommendation that cover different topics in computer science to facilitate research on code recommendation.
    \item We conduct extensive experiments on massive datasets to demonstrate the effectiveness of the proposed \model framework and its design choices.
\end{itemize}


\section{Preliminaries}

\subsection{GitHub}

GitHub is a code hosting platform for version control and collaboration based on \texttt{git}. 
Users can create \texttt{repositories}, namely, digital directories, to store source code for their projects. Users can make changes to the source code in the form of \texttt{commits}, which are snapshots of an OSS project, that capture the project state. A GitHub \texttt{commit} mainly reflects two aspects: 1) The commit author is highly interested in the project; 2) the user has the expertise to contribute.
In this work, we study the factors that influence users' contributing behaviors. 
We use git \texttt{commits} as positive interactions. 

\subsection{Problem Formulation}

Before delving into our proposed \model framework, we first formalize our code recommendation task. 
We use the terms ``repository'' and ``project'' interchangeably to refer to an open source project. 
We define $\mathcal{U}$, $\mathcal{V}$, $\mathcal{R}$ as the set of users, files, and repositories, respectively.
Each repository $r_k \in \mathcal{R}$ contains a subset of files $\mathcal{V}_k \subsetneq \mathcal{V}$. Both macroscopic project-level interactions and microscopic file-level interactions are present in OSS development.

\underline{File-level behaviors.} 
We define $\mathbf{Y} \in \{0, 1\}^{|\mathcal{U}| \times |\mathcal{V}|}$ as the interaction matrix between $\mathcal{U}$ and $\mathcal{V}$ for the file-level contribution behavior, where each entry is denoted by $y_{ij}$. $y_{ij} = 1$ indicates that $u_i$ has contributed to $v_j$, and $y_{ij} = 0$, otherwise.

\underline{Project-level behaviors.} 
Interactions at the project level are more diverse. For example, the popular code hosting platform GitHub allows users to \textit{star} (publicly bookmark) interesting repositories and \emph{watch} (subscribe to) repositories for updates. 
We thus define $\mathcal{T}$ as the set of user-project behaviors. 
Similar to $\mathbf{Y}$, we define $\mathbf{S}_{t} \in \{0, 1\}^{|\mathcal{U}| \times |\mathcal{R}|}$ as the project-level interaction matrix for behavior of type $t$.
Our goal is to predict the future file-level contribution behaviors of users based on their previous interactions. Formally, given the training data $\mathbf{Y}^{\textsc{tr}}$, we try to predict the interactions in the test set $y_{ij} \in \mathbf{Y}^{\textsc{ts}} = \mathbf{Y} \textbackslash \mathbf{Y}^{\textsc{tr}}$. 

\section{Methodology}

\begin{figure*}[t]
    \centering
    \includegraphics[width=0.92\linewidth]{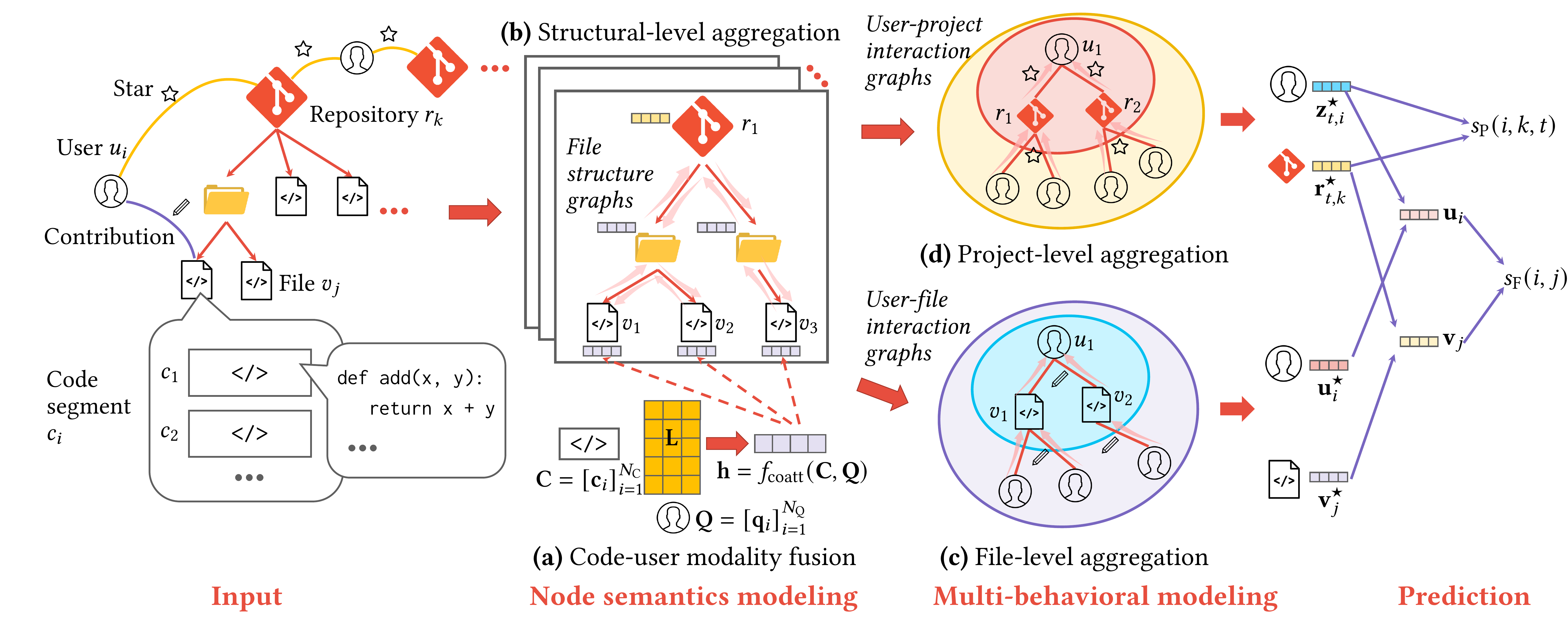}
    \caption{
        Our proposed \model framework for code recommendation. \model jointly considers project file structures, code semantics, and user behaviors. \model models the microscopic file-level interactions and macroscopic project-level interactions through Multi-Behavioral Modeling, and bridges the micro/macro-scopic signals through Node Semantics Modeling.
    }
    \label{fig:model}
\end{figure*}

As shown in Fig.~\ref{fig:model}, we design \model, a two-stage graph-based recommendation framework. \model considers $u_i \in \mathcal{U}, v_j \in \mathcal{V}, r_k \in \mathcal{R}$ as graph nodes, and models the user-item interactions and the item-item relations as edges. 
We use two sets of graphs to characterize the heterogeneous information in code recommendation. 
One is the user-item interaction graphs that encompass the collaborative signals. The other is the file-structure graphs that reveal file-file and file-project relationships from the project hierarchy perspective. 
The code recommendation problem is then formulated as a user-file link prediction task. 

\model contains two major components: 1) \emph{Node Semantics Modeling}, which learns the fine-grained representations of project files by fusing code semantics with their historical contributors, and then aggregate project hierarchical information on the file structure graph to learn the file and repository representation; 2) \emph{Multi-behavioral Modeling}, which jointly models the microscopic user-file interactions and macroscopic user-project interactions. Finally, \model fuses the representations from multiple behaviors for prediction. This way, node semantics modeling bridges the coarse-grained and fine-grained interaction signals on the item side. Therefore, \model efficiently characterizes intra-project and inter-project differences, eventually uncovering latent user and item features that explain the interactions $\mathbf{Y}$.

\subsection{Node Semantics Modeling}
\label{node_semantic_modeling}
Node semantics modeling aims to learn file and repository representation. The challenge is how to inherently combine the semantic features of each project file with its interacted users and the project hierarchy. 
To address this challenge, we first use a \textbf{code-user modality fusion} mechanism (Fig.~\ref{fig:model}a) to fuse the file content modality and the historical users at the code level. Then, we embed the fine-grained collaborative signals from user-file interactions into the file representations. Next, we employ \textbf{structural-level aggregation} (Fig.~\ref{fig:model}b), which explicitly models the project structures as hierarchical graphs to enrich the file/repository representation with structural information. 
This step produces representation for each file $v_j$ and repository $r_k$, which serve as the input for user behavior modeling in Sec.~\ref{sec:multi_behavioral_modeling}. 
\subsubsection{Code-User Modality Fusion}
\label{sec:code_user_modality_fusion}

A project file is characterized by diverse semantic features including multiple method declarations and invocations, which are useful for explaining why a contributor is interested in it. 
Inspired by the success of pretrained language models~\cite{kenton2019bert, liu2019roberta, jin2022prototypical}, we use pretrained \textsf{CodeBERT}~\cite{feng2020codebert}, a bimodal language model for programming languages, 
to encode the rich semantic features of each file. \textsf{CodeBERT} is shown to generalize well to programming languages not seen in the pretraining stage, making it suitable for our setting where project files are written in diverse programming languages. 
Here, a straightforward way is to directly encode each file into a per-file latent representation. 
Such an encoding scheme has two issues. 
Firstly, a file may contain multiple classes and function declarations that are semantically distinct. Fig.~\ref{fig:motivating_Example} shows the file structure of the \texttt{huggingface/transformers}~\cite{wolf2020transformers} repository as an example. The \texttt{modeling\_bert.py} file contains not only various implementations of the BERT language model for different NLP tasks, but also utilities for parameter loading and word embeddings. These implementations are distributed among several code segments in the same project file, and file-level encoding can fail to encode such semantic correlations. 
Secondly, the property of a project file is influenced by the historical contributors' features. A user's contribution can be viewed as injecting her/his own attributes, including programming style and domain knowledge, into the interacted file. Such contribution behaviors make it more likely to be interacted again by users with similar levels of expertise than random contributors.

Therefore, we propose a \textbf{code-user modality fusion} strategy to embed both code semantics and user characteristics into the file representation. Specifically, for each file, we partition its source code into $N_C$ code segments and encode each of them into a code-segment-level representation $\mathbf{c}_i$. This produces a feature map $\mathbf{C} = [\mathbf{c}_1, \mathbf{c}_2, \ldots \mathbf{c}_{N_\text{C}}], \mathbf{C} \in \mathbb{R}^{N_\text{C} \times d}$, where $d$ is the embedding size. Similarly, we sample $N_Q$ historical users of the file and encode them into a feature map $\mathbf{Q} = [\mathbf{u}_1, \mathbf{u}_2, \ldots \mathbf{u}_{N_\text{Q}}], \mathbf{Q} \in \mathbb{R}^{N_\text{Q} \times d}$. Please refer to Appendix~\ref{supp:implementation} for details in encoding $\mathbf{C}$ and $\mathbf{Q}$. Inspired by the success of co-attention~\cite{yu2019deep, lu2016hierarchical}, we transform the user attention space to code attention space by calculating a code-user affinity matrix $\mathbf{L} \in \mathbb{R}^{N_\text{C} \times N_\text{Q}}$:
\begin{equation}
\mathbf{L}=\tanh (\mathbf{C} \mathbf{W}_\text{O} \mathbf{Q}^{\top}),
\label{eq:coattention_affinity_matrix}
\end{equation}
where $\mathbf{W}_\text{O} \in \mathbb{R}^{d \times d}$ is a trainable weight matrix. Next, we compute the attention weight $\mathbf{a} \in \mathbb{R}^{N_\text{C}}$ of the code segments to select salient features from $\mathbf{C}$. 
We treat the affinity matrix as a feature and learn an attention map $\mathbf{H}$ with $N_\text{H}$ representations:
\begin{align}
\mathbf{H}&=\tanh (\mathbf{W}_\text{C} \mathbf{C}^{\top}+\mathbf{W}_\text{Q} (\mathbf{L}\mathbf{Q})^{\top} ), \\
\mathbf{a}&=\operatorname{softmax}(\mathbf{w}_\text{H}^{\top} \mathbf{H}),
\end{align}
where $\mathbf{W}_\text{C}, \mathbf{W}_\text{Q} \in \mathbb{R}^{N_\text{H} \times d}, \mathbf{w}_\text{H} \in \mathbb{R}^{N_\text{H}}$ are the weight parameters. Finally, the file attention representation $\mathbf{h}$ is calculated as the weighted sum of the code feature map:
\begin{equation}
\mathbf{h} = \mathbf{a}^{\top} \mathbf{C}. 
\end{equation}
The file attention representation serves as a start point to further aggregate file structural features. 

\subsubsection{Structural-Level Aggregation}
\label{sec:structure_level_aggregation}

Projects are organized in a hierarchical way such that nodes located closer on the file structure graph are more closely related in terms of semantics and functionality. 
For example, in Fig.~\ref{fig:motivating_Example}, both files under the \path{bert/} directory contain source code for the BERT~\cite{kenton2019bert} language model, and files under \path{roberta/} contains implementation for the RoBERTa~\cite{liu2019roberta} model. The file \path{modeling_bert.py} is therefore more closely related to \path{tokenization_bert.py} in functionality than to \path{tokenization_roberta.py}.

To exploit such structural clues, we model each repository as a hierarchical heterogeneous graph~\cite{yu2021graph} $G_\text{S}$ consisting of file, directory, and repository nodes. Each node is connected to its parent node through an edge, and nodes at the first level are directly connected to the virtual root node representing the project. 
To encode the features of directory nodes, we 
partition the directory names into meaningful words according to underscores and letter capitalization, then encoded the nodes by their TF-IDF features. Our encoding scheme is motivated by the insight that the use of standard directory names (e.g., \texttt{doc}, \texttt{test}, and \texttt{models}) is correlated with project popularity among certain groups of developers~\cite{zhu2014patterns, borges2016understanding}. Repository features are encoded by their project owners, creation timestamps, and their top-5 programming languages. 
The repository and directory representations are mapped to the same latent space as the file nodes. 
Then, we pass the representation $\mathbf{h}$ through multiple GNN layers to aggregate the feature of each node from its neighbors on $G_\text{S}$:
\begin{equation}
    \widetilde{\mathbf{h}} = f_{\text{GNN}}(\mathbf{h}, G_\text{S}),
    \label{eq:gnn}
\end{equation}
where $\widetilde{\mathbf{h}}$ is the structure-enhanced node representation. The aggregation function $ f_{\text{GNN}}(\cdot)$ can be chosen from a wide range of GNN architectures, such as GCN~\cite{kipf2016semi}, GraphSAGE~\cite{hamilton2017inductive}, and GIN~\cite{xu2018powerful}. 
In practice, we employ a 3-layer Graph Attention Network (GAT)~\cite{velivckovic2018graph}. 
\subsection{Multi-behavioral Modeling}
\label{sec:multi_behavioral_modeling}
Direct modeling of the sparse contribution behavior potentially leads to inaccurate user/item representations and aggravates the cold-start issue.
Instead, we jointly model the microscopic user-file contribution in File-level Aggregation (Fig.~\ref{fig:model}c) and macroscopic user-project interactions in Project-level Aggregation (Fig.~\ref{fig:model}d) to learn user preferences and address the sparsity issue. Then, the representations learned from multi-level behaviors are combined to form the user and item representations for prediction. 

\subsubsection{File-level Aggregation}
\label{sec:intra_level_aggregation}

We model the project files and their contributors as an undirected user-file bipartite graph $\mathcal{G}_\text{F}$ consisting of users $u_i \in \mathcal{U}$, files $v_j \in \mathcal{V}$ and their interactions. The initial embedding matrix of users/items is denoted by $\mathbf{E}^{(0)}$:
\begin{equation}
\mathbf{E}^{(0)}=\bigl[\underbrace{\mathbf{u}_{1}^{(0)}, \cdots, \mathbf{u}_{|\mathcal{U}|}^{(0)}}_{\text {users embeddings}}, \underbrace{\mathbf{v}_{1}^{(0)}, \cdots, \mathbf{v}_{|\mathcal{V}|}^{(0)}}_{\text {item embeddings}}\bigr],
\label{eq:initial_embedding_file_level}
\end{equation}
where $\mathbf{u}_{i}^{(0)}$ is the initial embedding for user $u_i$ and $\mathbf{v}_{j}^{(0)}$ is the initial embedding for file $v_j$, equivalent to its structure-enhanced representation $\widetilde{\mathbf{h}}$ 
(Sec.~\ref{sec:structure_level_aggregation}). We adopt the simple weight sum aggregator in LightGCN~\cite{he2020lightgcn} in the propagation rule: 
\begin{align}
\mathbf{u}_i^{(l)}&=\sum_{v_j \in \mathcal{N}_{i}} \frac{1}{\sqrt{\left|\mathcal{N}_i\right|\left|\mathcal{N}_j\right|}}  \mathbf{v}_j^{(l-1)}, \\ 
\mathbf{v}_j^{(l)}&=\sum_{u_i \in \mathcal{N}_{j}} \frac{1}{\sqrt{\left|\mathcal{N}_j\right|\left|\mathcal{N}_i\right|}} \mathbf{u}_i^{(l-1)},
\end{align}
where $\mathbf{u}_{i}^{(l)}$ and $\mathbf{v}_{j}^{(l)}$ are the embeddings for user $u_i$ and file $v_j$ at layer $l$, $\mathcal{N}_{i}$ and $\mathcal{N}_{j}$ indicate the neighbors of user $u_i$ and file $v_j$, and $1 / \sqrt{|\mathcal{N}_i||\mathcal{N}_j|}$ is the symmetric normalization term set to the graph Laplacian norm to avoid the increase of GCN embedding sizes~\cite{kipf2016semi, he2020lightgcn}. In the matrix form, the propagation rule of file-level aggregation can be expressed as: 
\begin{align}
\mathbf{E}^{(l)} = \mathbf{D}^{-\frac{1}{2}}\mathbf{A}\mathbf{D}^{-\frac{1}{2}} \mathbf{E}^{(l-1)},
\label{fig:affinity_matrix_contrib}
\end{align}
where $\mathbf{A} =\left(\begin{array}{cc}
\mathbf{0} & \mathbf{Y}^{\textsc{tr}} \\
(\mathbf{Y}^{\textsc{tr}})^{\top} & \mathbf{0}
\end{array}\right) \in \mathbb{R}^{(|\mathcal{U}| + |\mathcal{V}|) \times (|\mathcal{U}| + |\mathcal{V}|)}$ is the affinity matrix, and 
$\mathbf{D}$ is the diagonal degree matrix in which each entry $\mathbf{D}_{ii}$ indicates the number of non-zero entries on the $i$-th row of $\mathbf{A}$. 
By stacking multiple layers, each user/item node aggregates information from its higher-order neighbors. Propagation through $L$ layers yields a set of representations $\{\mathbf{E}^{(l)}\}_{l=0}^{L}$. Each $\mathbf{E}^{(l)}$ emphasizes the messages from its $l$-hop neighbors. We apply mean-pooling over all $\mathbf{E}^{(l)}$ to derive the user and file representations $\mathbf{u}_i^{\star}$ and $\mathbf{v}_j^{\star}$ from different levels of user/item features:
\begin{align}
\mathbf{u}_i^{\star} &= \frac{1}{L+1} \sum_{l = 0}^{L} \mathbf{u}_i^{(l)}, \\
\mathbf{v}_j^{\star} &= \frac{1}{L+1} \sum_{l = 0}^{L} \mathbf{v}_j^{(l)}.
\end{align}

\subsubsection{Project-Level Aggregation}
\label{sec:inter_level_aggregation}
OSS development is characterized by both microscopic contribution behaviors and multiple types of macroscopic project-level behaviors. 
For example, developers usually find relevant projects and reuse their functions and explore ideas of possible features~\cite{husain2019codesearchnet, he2021pyart}.
In particular, GitHub users can \emph{star} (bookmark) 
interesting repositories and discover projects under similar topics. 
This way, developers can adapt code implementation of these interesting projects into their own development later. 
Hence, project-level macroscopic interactions are conducive for extracting users' broad interests. 

For each behavior $t$, we propagate the user and repository embeddings on its project-level interaction graph $\mathcal{G}_\text{P}^t$:
\begin{equation}
    \mathbf{Z}_{t}^{l} = \mathbf{D}_{t}^{-\frac{1}{2}}\mathbf{\Lambda}_{t}\mathbf{D}_{t}^{-\frac{1}{2}} \mathbf{Z}_{t}^{(l-1)}.
\label{eq:gcn_project_level}
\end{equation}
The initial embeddings $\mathbf{Z}^{(0)}$ is shared by all $t \in \mathcal{T}$ and is composed of the initial user representations identical to Eq.~(\ref{eq:initial_embedding_file_level}) and the repository embeddings from the structure-enhanced representation $\widetilde{\mathbf{h}}$ in Eq.~(\ref{eq:gnn}): 
\begin{align}
\mathbf{Z}^{(0)} &= \bigl[\underbrace{\mathbf{z}_1^{(0)}, \mathbf{z}_2^{(0)}, \ldots \mathbf{z}_{|\mathcal{U}|}^{(0)}}_{\text{user embeddings}},\underbrace{\mathbf{r}_1^{(0)}, \mathbf{r}_2^{(0)}, \ldots \mathbf{r}_{|\mathcal{R}|}^{(0)}}_{\text{repository embeddings}}\bigr], \label{eq:initial_embedding_repo_level}
\end{align}
where $\mathbf{z}_i^{(0)} = \mathbf{u}_i^{(0)}$, $\mathbf{\Lambda}_t \in \mathbb{R}^{(|\mathcal{U}| + |\mathcal{R}|) \times (|\mathcal{U}| + |\mathcal{R}|)}$ is the affinity matrix for behavior $t$ constructed similarly as $\mathbf{A}$ in Eq.~(\ref{fig:affinity_matrix_contrib}). With representations $\{\mathbf{Z}_{t}^{(l)}\}_{l=0}^{L}$ obtained from multiple layers, we derive the combined user and repository representations for behavior $t$ as
\begin{align}
\mathbf{z}_{t, i}^{\star} &= \frac{1}{L+1} \sum_{l = 0}^{L} \mathbf{z}_{t, i}^{(l)}, \\
\mathbf{r}_{t, i}^{\star} &= \frac{1}{L+1} \sum_{l = 0}^{L} \mathbf{r}_{t, i}^{(l)}.
\end{align}

\subsection{Prediction}

For file-level prediction, we aggregate the macroscopic signals $\mathbf{z}_{t, i}^{\star}, \mathbf{r}_{t, i}^{\star}$ from each behavior $t$ into $\mathbf{u}_i, \mathbf{v}_j$: 
\begin{align}
\mathbf{z}_i^{\star} & = \textsc{AGG}(\{z_t^{\star}, t \in \mathcal{T}\}), &\quad
\mathbf{r}_k^{\star} & = \textsc{AGG}(\{r_t^{\star}, t \in \mathcal{T}\}), \\
\mathbf{u}_i &= \textsc{MLP}([\mathbf{u}_i^{\star} \mathbin\Vert \mathbf{z}_i^{\star}]), &\quad
\mathbf{v}_j &= \textsc{MLP}( [\mathbf{v}_j^{\star} \mathbin\Vert \mathbf{r}_{\phi({j})}^{\star}]),
\end{align}
where $\textsc{AGG}(\cdot)$ 
is an aggregation function and $\textrm{MLP}(\cdot)$ is a multilayer perceptron, 
$\phi(\cdot): \mathcal{V} \rightarrow \mathcal{R}$ maps the index of each project file to its repository, and $\mathbin\Vert$ is the concatenation operator. 
On the user side, both macroscopic interests and micro-level interactions are injected into the user representations. On the item side, the semantics of each file is enriched by its interacted users and the repository structural information.

For computational efficiency, we employ inner product to calculate the user $u_i$'s preference towards each file $v_j$:
\begin{equation}
s_\text{F}(i, j) = \mathbf{u}_i^{\top} \mathbf{v}_j,
\label{eq:file_level_prediction}
\end{equation}
where $s_\text{F}$ is the scoring function for the file-level behavior. 
Similarly, for each user-project pair, we derive a project-level score for each behavior $t$ using the project-level scoring function $s_\text{P}$: 
\begin{equation}
s_\text{P}(i, k, t) = (\mathbf{z}_{t, i}^{\star})^\top \mathbf{r}_{t, k}^{\star}.
\label{eq:repo_level_prediction}
\end{equation}

\subsection{Optimization}
\label{sec:optimization}

We employ the Bayesian Personalized Ranking (BPR)~\cite{rendle2009bpr} loss, which encourages the prediction of an observed user-item interaction to be greater than an unobserved one:
\begin{align}
\mathcal{L}_\text{F} &=\sum_{(i, j^{+}, j^{-}) \in \mathcal{O}}-\log (\operatorname{sigmoid}(s_\text{F}(i, j^{+}) - s_\text{F}(i, j^{-}))), \\
\mathcal{L}_\text{P}^{t} &=\sum_{(i, k^{+}, k^{-}) \in \mathcal{O}}-\log(\operatorname{sigmoid}(s_\text{P}(i, k^{+}, t) - s_\text{P}(i, k^{-}, t))),
\label{eq:bpr_loss}
\end{align}
where $\mathcal{L}_\text{F}$ is the file-level BPR loss, and $\mathcal{L}_\text{P}^{t}$ is the project-level BPR loss for behavior $t$, $\mathcal{O}$ denotes the pairwise training data, $j^{+}$ indicates an observed interaction between user $u_i$, and item $v_{j^{+}}$ and $j^{-}$ indicates an unobserved one. As high-order neighboring relations within contributors are also useful for recommendations, we enforce users to have similar representations as their structural neighbors through structure-contrastive learning objective~\cite{lin2022improving, wu2022adversarial}: 
\begin{equation}
\mathcal{L}_\text{C}^\mathcal{U}=\sum_{u_i \in \mathcal{U}}-\log \frac{\exp (\mathbf{u}_i^{(\eta)} \cdot \mathbf{u}_i^{(0)} / \tau)}{\sum_{u_j \in \mathcal{U}} \exp (\mathbf{u}_i^{(\eta)} \cdot \mathbf{u}_j^{(0)} / \tau)}.
\label{eq:structure_loss_user}
\end{equation}
Here, $\eta$ is set to an even number so that each user node can aggregate signals from other user nodes and $\tau$ is a temperature hyper-parameter. Similarly, the contrastive loss is applied to each $v_i$: 
\begin{equation}
\mathcal{L}_\text{C}^\mathcal{V}=\sum_{v_i \in \mathcal{V}}-\log \frac{\exp (\mathbf{v}_i^{(l)} \cdot \mathrm{v}_i^{(0)} / \tau)}{\sum_{v_j \in \mathcal{V}} \exp (\mathbf{v}_i^{(l)} \cdot \mathbf{v}_j^{(0)} / \tau)}.
\label{eq:structure_loss_file}
\end{equation}
The overall optimization objective is 
\begin{equation}
    \mathcal{L} = \mathcal{L}_\text{F} + \lambda_1 \sum_{t \in \mathcal{T}} \mathcal{L}_\text{P}^t + \lambda_2 (\mathcal{L}_\text{C}^\mathcal{U} + \mathcal{L}_\text{C}^\mathcal{V}) + \lambda_3 \|\Theta\|_2,
\end{equation}
where $\Theta$ denotes all trainable model parameters and $\lambda_1, \lambda_2, \lambda_3$ are hyper-parameters.

\subsection{Complexity Analysis}
The node semantic modeling (Sec.~\ref{node_semantic_modeling}) has time complexity of $O((N_\text{C} + N_\text{Q})|\mathcal{V}| + |\mathcal{E}|)$, where $\mathcal{E}$ is the set of edges in all file structure graphs.
The user behavior modeling (Sec.~\ref{sec:multi_behavioral_modeling}) has time complexity of $O(|\mathbf{A}^{+}| + \sum_{t} |\mathbf{A}_{t}^{+}|)$, where $|\mathbf{A}^{+}|$ is the number of positive entries in $|\mathbf{A}|$ and likewise for $|\mathbf{\Lambda}_{t}^{+}|$.
The overall time complexity is $O((N_\text{C} + N_\text{Q})|\mathcal{V}| + |\mathcal{E}| + |\mathbf{A}^{+}| +  \sum_{t} |\mathbf{\Lambda}_{t}^{+}|)$. Although obtaining the initial code segment embeddings $\mathbf{C}$ implies large computational costs, our model only calculates $\mathbf{C}$ once and caches it to be used in each iteration.
Empirically, the average inference time of MF, LightGCN, and \model are 0.804 ms, 1.073 ms, and 1.138 ms per test example, respectively.

\section{Experiments}

\subsection{Experimental Settings}

\subsubsection{Datasets}

We collected 3 datasets covering diverse topics in computer science including machine learning (ML), fullstack (FS), and database (DB), using the GitHub API~\footnote{\url{https://docs.github.com/en/rest}} and the PyGithub~\footnote{\url{https://github.com/PyGithub/PyGithub.git}}package.  
We retain repositories with $\ge 250$ stars and $\ge 3$ contributors to exclude repositories intended for private usages~\cite{borges2016understanding}. We include projects with contribution history of at least 3 months according to their commit history. 
To ensure that our model generalizes on a wide range of topics, popularity, and project scales, we first select 3 subsets of repositories using their GitHub topics~\footnote{\url{https://github.com/topics}}, which are project labels created by the project owners. Then, we randomly sample 300 repositories from each subset considering their numbers of project files and stars. 
We use the Unix timestamp \texttt{1550000000} and \texttt{1602000000} to partition the datasets into training, validation, and test sets. This way, all interactions before the timestamp are used as the training data. We retain the users with at least 1 interaction in both train and test set. More details about dataset construction are in the appendix.
\begin{table}
\centering
\caption{Summary of the datasets. The second column shows the number of files with observed interactions instead of all existing files in the projects.}
\begin{tabular}{c|cccc} 
\hline
\hline
Dataset & \#Files & \#Users & \#Interactions & Density  \\ 
\hline
ML      & 239,232  & 21,913       & 663,046  & $1.26 \times 10^{-4}$  \\
DB      & 415,154  & 30,185       & 1,935,155 & $1.54\times 10^{-4}$  \\
FS      & 568,972  & 51,664       & 1,512,809 & $5.14\times 10^{-5}$  \\
\hline
\hline
\end{tabular}
\label{fig:dataset}
\end{table}

\begin{table*}
\centering
\caption{The overall performance on 3 datasets. The best performance is marked in bold. The second best is underlined.}
results among
\setlength{\tabcolsep}{3.1mm}
\begin{tabular}{l|l|ccccccc|c} 
\hline
\hline
 Dataset & Metric  & MF   & MLP   & NeuMF   & NGCF  & LightGCN & NCL   &   \model & Improvement   \\ 
\hline
        & NDCG@5  & 0.065                   & 0.073                   & 0.076                   & 0.091                    & 0.106                   & \underline{0.119}           & \textbf{0.132}             & 11.2\%                     \\
        & Hit@5   & 0.162                   & 0.189                   & 0.189                   & 0.237                    & \underline{0.291}           & 0.276                   & \textbf{0.351}             & 20.5\%                     \\
        & MRR@5   & 0.098                   & 0.113                   & 0.114                   & 0.137                    & 0.164                   & \underline{0.201}           & \textbf{0.211}             & 5.0\%                      \\
        & NDCG@10 & 0.066                   & 0.075                   & 0.081                   & 0.093                    & 0.109                   & \underline{0.118}           & \textbf{0.136}             & 14.8\%                     \\
ML      & Hit@10  & 0.229                   & 0.250                   & 0.263                   & 0.310                    & \underline{0.386}           & 0.337                   & \textbf{0.440}             & 14.0\%                     \\
        & MRR@10  & 0.106                   & 0.121                   & 0.124                   & 0.147                    & 0.177                   & \underline{0.209}           & \textbf{0.223}             & 6.7\%                      \\
        & NDCG@20 & 0.072                   & 0.081                   & 0.084                   & 0.100                    & 0.116                   & \underline{0.120}           & \textbf{0.141}             & 17.9\%                     \\
        & Hit@20  & 0.324                   & 0.343                   & 0.346                   & 0.407                    & 0.457                   & \underline{0.466}           & \textbf{0.540}             & 15.8\%                     \\
        & MRR@20  & 0.113                   & 0.127                   & 0.130                   & 0.154                    & 0.185                   & \underline{0.213}           & \textbf{0.230}             & 8.2\%                      \\ 
\hline
        & NDCG@5  & 0.085                   & 0.079                   & 0.085                   & 0.099                    & 0.082                   & \underline{0.124}           & \textbf{0.160}             & 29.0\%                     \\
        & Hit@5   & 0.205                   & 0.191                   & 0.206                   & 0.263                    & 0.237                   & \underline{0.316}           & \textbf{0.390}             & 23.2\%                     \\
        & MRR@5   & 0.130                   & 0.118                   & 0.128                   & 0.162                    & 0.132                   & \underline{0.252}           & \textbf{0.260}             & 3.2\%                      \\
        & NDCG@10 & 0.086                   & 0.079                   & 0.085                   & 0.100                    & 0.084                   & \underline{0.123}           & \textbf{0.159}             & 29.4\%                     \\
DB      & Hit@10  & 0.267                   & 0.251                   & 0.276                   & 0.361                    & 0.324                   & \underline{0.380}           & \textbf{0.488}             & 28.4\%                     \\
        & MRR@10  & 0.138                   & 0.126                   & 0.137                   & 0.175                    & 0.144                   & \underline{0.260}           & \textbf{0.273}             & 4.9\%                      \\
        & NDCG@20 & 0.088                   & 0.083                   & 0.088                   & 0.103                    & 0.091                   & \underline{0.125}           & \textbf{0.160}             & 27.3\%                     \\
        & Hit@20  & 0.335                   & 0.338                   & 0.362                   & 0.454                    & 0.422                   & \underline{0.437}           & \textbf{0.588}             & 29.5\%                     \\
        & MRR@20  & 0.143                   & 0.132                   & 0.143                   & 0.182                    & 0.150                   & \underline{0.264}           & \textbf{0.280}             & 6.0\%                      \\ 
\hline
        & NDCG@5  & 0.063                   & 0.063                   & 0.067                   & 0.082                    & 0.089                   & \underline{0.106}           & \textbf{0.146}             & 37.1\%                     \\
        & Hit@5   & 0.168                   & 0.178                   & 0.179                   & 0.231                    & 0.245                   & \underline{0.283}           & \textbf{0.374}             & 31.9\%                     \\
        & MRR@5   & 0.100                   & 0.100                   & 0.107                   & 0.132                    & 0.146                   & \underline{0.170}           & \textbf{0.226}             & 33.0\%                     \\
        & NDCG@10 & 0.063                   & 0.065                   & 0.068                   & 0.085                    & 0.092                   & \underline{0.106}           & \textbf{0.144}             & 35.6\%                     \\
FS      & Hit@10  & 0.231                   & 0.244                   & 0.249                   & 0.319                    & 0.332                   & \underline{0.361}           & \textbf{0.467}             & 29.3\%                     \\
        & MRR@10  & 0.109                   & 0.110                   & 0.117                   & 0.144                    & 0.157                   & \underline{0.180}           & \textbf{0.239}             & 32.3\%                     \\
        & NDCG@20 & 0.067                   & 0.070                   & 0.073                   & 0.090                    & 0.095                   & \underline{0.110}           & \textbf{0.146}             & 32.7\%                     \\
        & Hit@20  & 0.307                   & 0.321                   & 0.335                   & 0.406                    & 0.414                   & \underline{0.451}           & \textbf{0.559}             & 23.9\%                     \\
        & MRR@20  & 0.114                   & 0.115                   & 0.122                   & 0.150                    & 0.163                   & \underline{0.187}           & \textbf{0.245}             & 31.4\%                     \\
\hline
\hline
\end{tabular}
\label{tab:same_repo}
\end{table*}

\subsubsection{Implementation Details}
We implemented our \model model in PyTorch~\cite{paszke2019pytorch} and PyG~\cite{Fey/Lenssen/2019}. For all models, we set the embedding size to 32 and perform Xavier initialization~\cite{glorot2010understanding} on the model parameters. We use Adam optimizer~\cite{kingma2015adam} with a batch size of 1024. 
For Node Semantic Modeling (Sec.~\ref{node_semantic_modeling}), we conduct a grid search on $N_\text{C} \in \{4, 8, 12, 16\}$ and $N_\text{Q} \in \{2, 4, 8\}$, and choose to set $N_\text{C}=8$ and $N_\text{Q}=4$. 
The code encoder we use is the pretrained CodeBERT~\cite{feng2020codebert} model with 6 layers, 12 attention heads, and 768-dimensional hidden states.
For Multi-Behavioral Modeling (Sec.~\ref{sec:multi_behavioral_modeling}), we set the number of convolution layers $L=4$ for both intra- and inter-level aggregation. 
For prediction and optimization, we search the hyper-parameter $\lambda_3$ in $\{10^{-4}, 10^{-3}, 10^{-2}\}$, and $\lambda_1$ in~$\{10^{-2}, 10^{-1}, 1\}$. For the structure contrastive loss~\cite{lin2022improving}, we adopt the hyper-parameter setting from the original implementation and set $\lambda_2=10^{-6}, \eta=2$ without further tuning. 
For the baseline models, the hyper-parameters are set to the optimal settings as reported in their original papers. For all models, we search the learning rate in~$\{10^{-4}, 3\times 10^{-4}, 10^{-3}, 3\times 10^{-3}, 10^{-2}\}$.

\subsubsection{Baselines}
We compare \model with 3 groups of methods:
\begin{itemize}[wide]
    \item \textbf{G1}: a factorization-based method MF~\cite{rendle2009bpr},
    \item \textbf{G2}: neural-network-based methods including MLP~\cite{srivastava2012multimodal} and Neu-MF~\cite{he2017neural},
    \item \textbf{G3}: graph-based methods that model user-item interactions as graphs, including NGCF~\cite{wang2019neural}, LightGCN~\cite{he2020lightgcn}, and NCL~\cite{lin2022improving}.
\end{itemize}
As the task is to predict users' file-level contribution, file-level behavior modeling is the most critical component. Therefore, we use file-level contribution behaviors as the supervision signals as in Eq.~(\ref{eq:file_level_prediction}). For brevity, we use \underline{repository identity} to refer to the information of which repository a file belongs to. 
As the baselines do not explicitly leverage the repository identities of files, we encode their repository identities as a categorical feature through one-hot encoding during embedding construction. 
To ensure fairness of comparison, we incorporate the project-level interaction signals into the user representations by applying multi-hot encoding on the repositories each user has interacted with. 
All the baseline models use the same pretrained CodeBERT embeddings as \model to leverage the rich semantic features in the source code. 

\subsubsection{Evaluation Metrics}
Following previous works~\cite{he2020lightgcn, wang2019neural, he2017neural, zheng2021disentangling}, we choose Mean Reciprocal Rank (MRR@$K$), Normalized Discounted Cumulative Gain (NDCG@$K$), Recall@$K$ (Rec@$K$), and Hit@$K$ as the evaluation metrics.


\subsection{Performance}

\subsubsection{Intra-Project Recommendation}
\label{sec:overall_performances}

In this setting, we evaluate the model's ability to recommend development tasks under her interacted repositories.
For each user $u_i$, we rank the interactions under repositories they have interacted with in the training set. 
This setting corresponds to the scenario in which project maintainers recommend new development tasks to existing contributors based on their previous contribution. 
As shown in Tab.~\ref{tab:same_repo}, \model consistently outperforms the baselines by a large margin. On the ML dataset, \model outperforms the best baseline by 17.9\% on NDCG@20, 15.8\% on Hit@20, and 8.2\% on MRR@20. On the DB dataset, \model achieves performance improvements of 27.3\% on NDCG@20, 29.5\% on Hit@20, and 6.0\% on MRR@20. Notably, the greatest performance improvement is achieved on the FS dataset, which has the greatest sparsity. \model achieves a maximum performance improvement of 37.1\% on NDCG@5 and 35.6\% on NDCG@10. The results show that \model achieves significant performance improvement over the baseline, and is especially useful when the observed interactions are scarce. 

Among the baselines, graph-based method (\textbf{G3}) achieves better performances than (\textbf{G1}), (\textbf{G2}) as they can model the high-order relations between users and items through the interaction graph and the embedding function. LightGCN~\cite{he2020lightgcn} underperforms NGCF~\cite{wang2019neural} on the DB dataset, whose training set has the greatest density, and outperforms NGCF on the ML and FS datasets. This implies that the message passing scheme of NGCF, which involves multiple linear transformation and non-linear activation, is more effective for denser interactions. Such results justify our design choice in multi-behavioral modeling, which uses the LightGCN propagation scheme. 
NCL exhibits the strongest performance, demonstrating the importance of the contrastive learning loss in modeling differences among homogeneous types of nodes, which is also included in our model design. Neural-network-based methods (\textbf{G2}) generally outperform matrix factorization (\textbf{G1}), as they leverage multiple feature transformations to learn the rich semantics in the file embeddings and the user-file interactions. 

\begin{figure}
  \centering
  \includegraphics[width=0.45\linewidth]{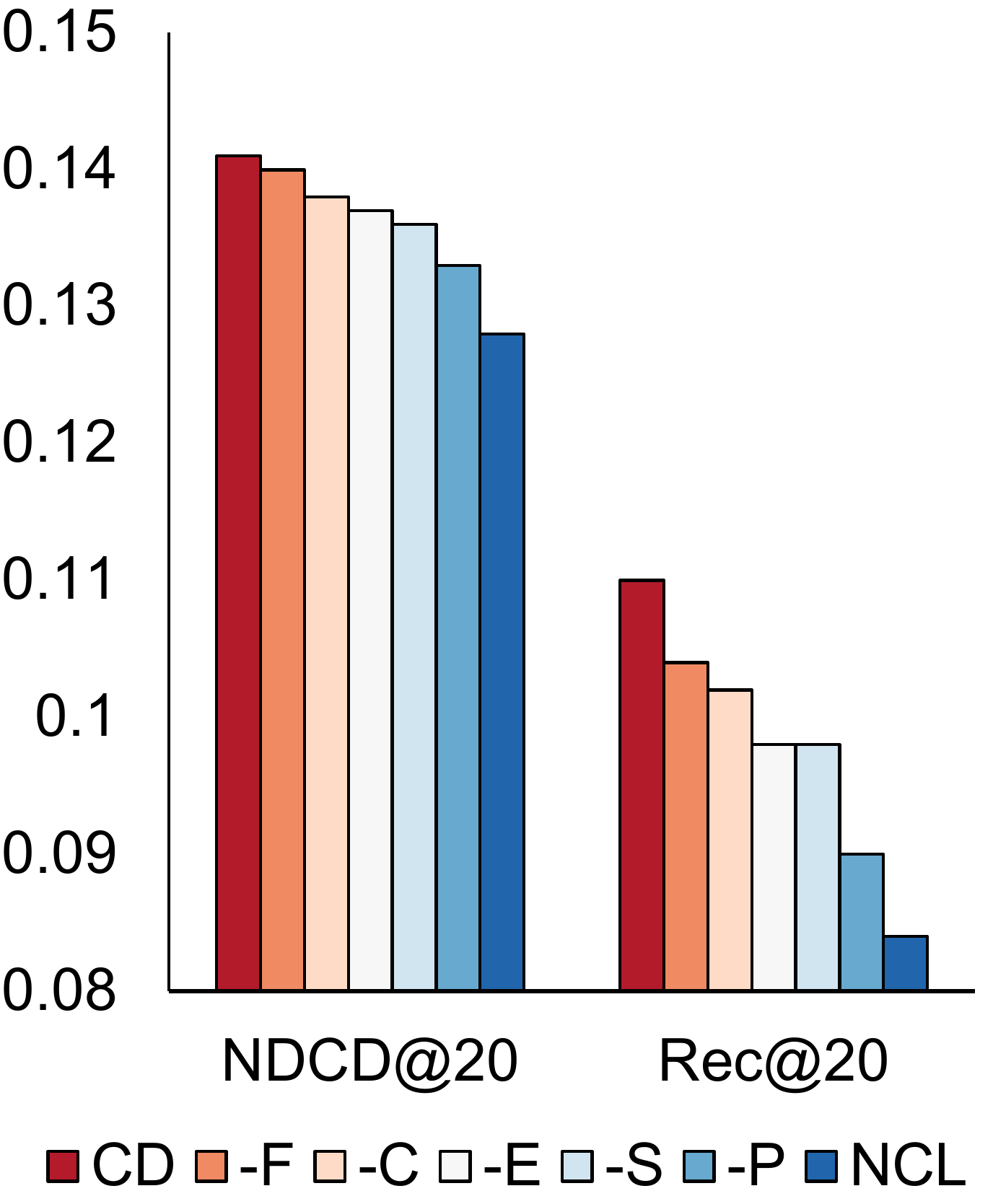}\quad
  \includegraphics[width=0.45\linewidth]{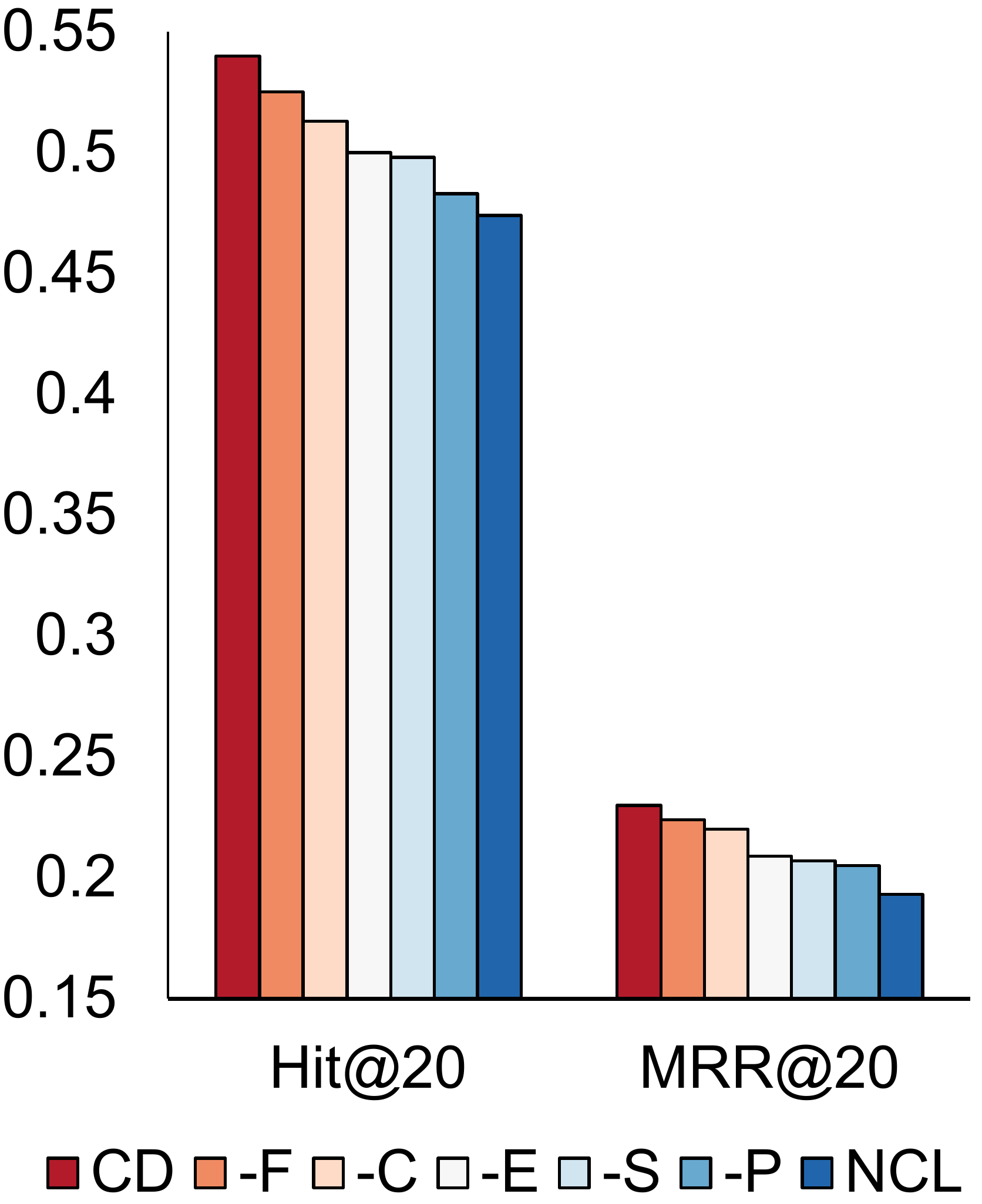}
  \caption{
  Results among variants of \model and the best baseline model NCL on the ML dataset. 
  }
  \label{fig:ablation_ml}
\end{figure}

\begin{figure}
  \centering
  \includegraphics[width=0.45\linewidth]{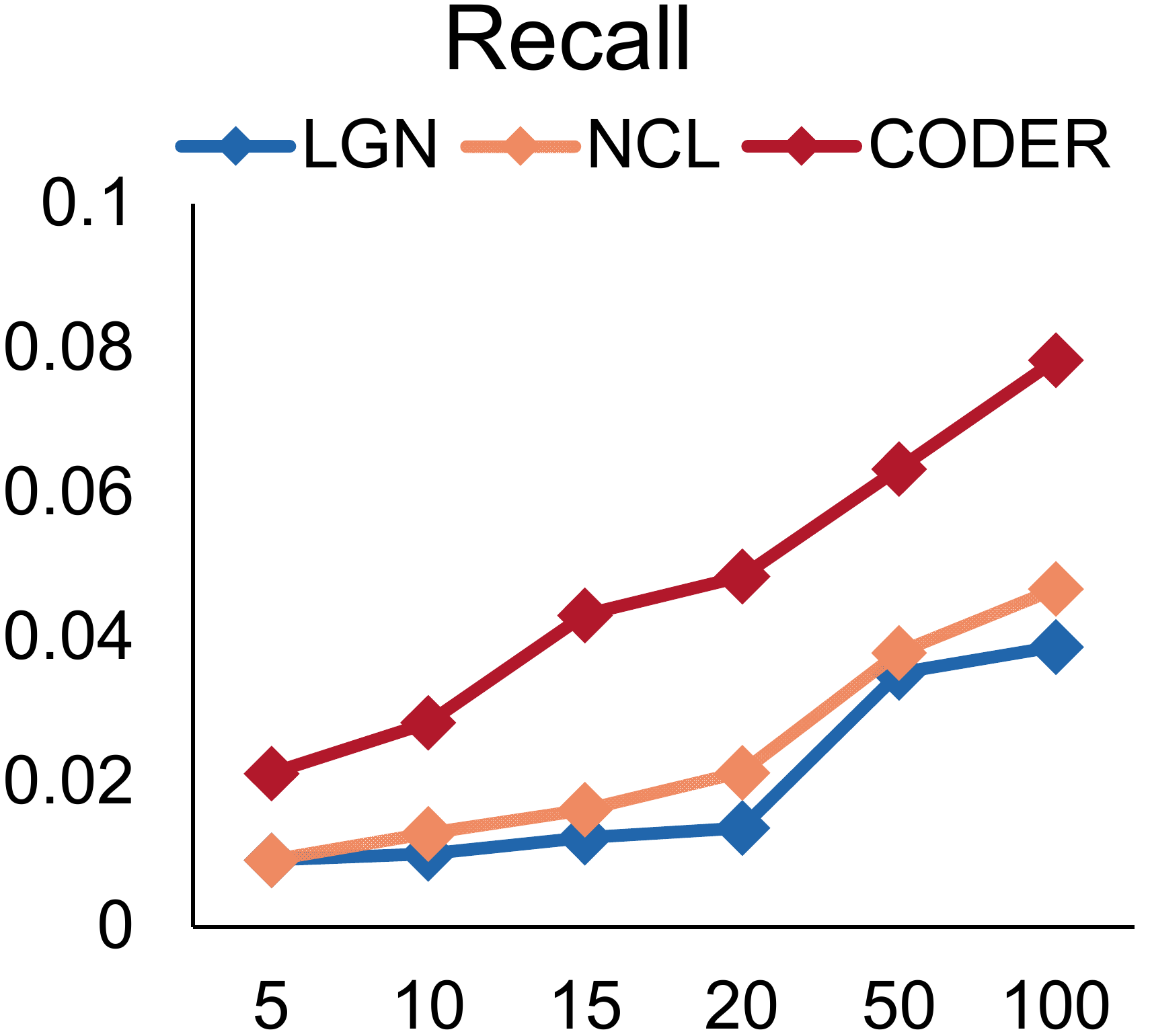}\quad
  \includegraphics[width=0.45\linewidth]{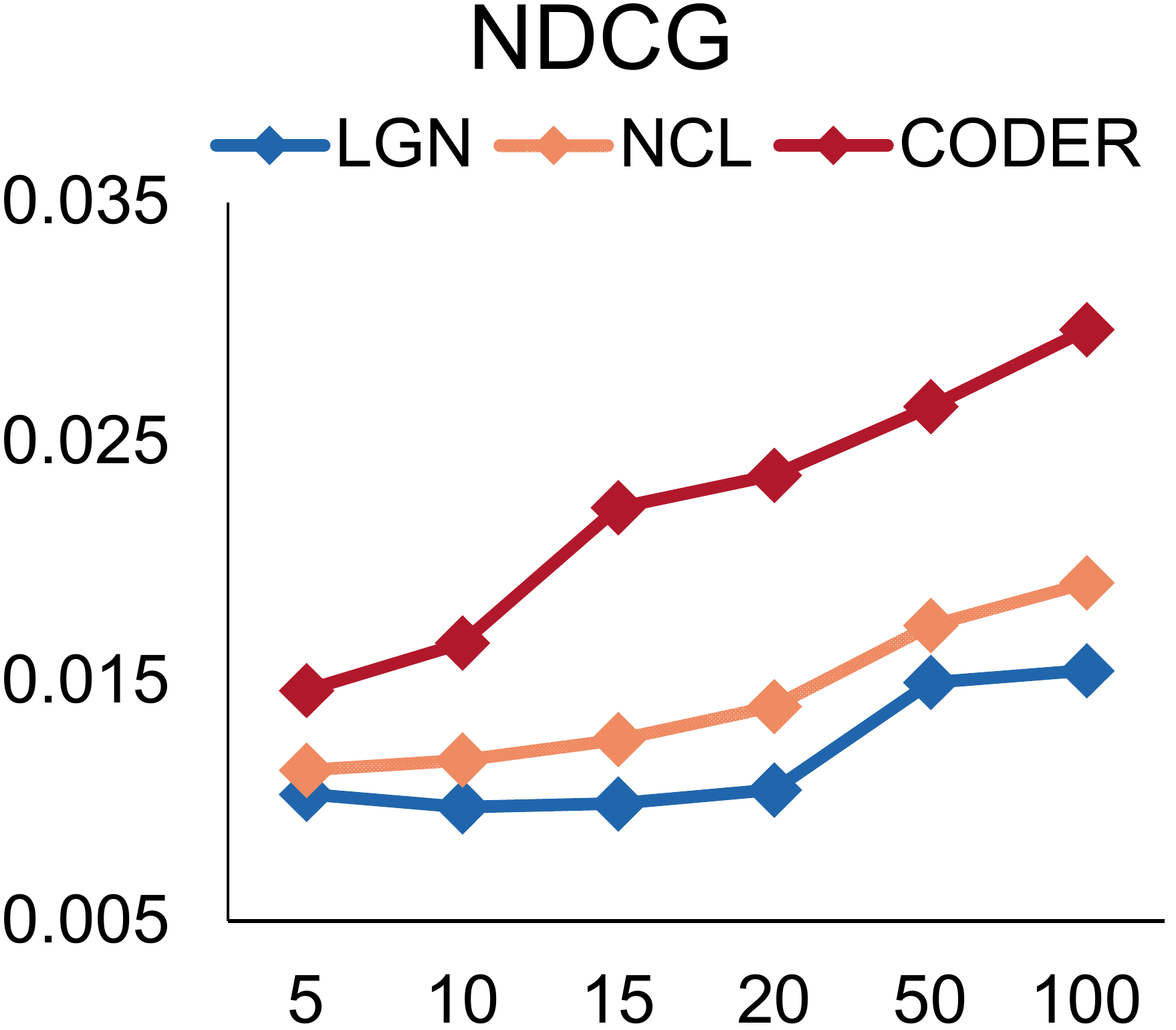}\\
  \includegraphics[width=0.45\linewidth]{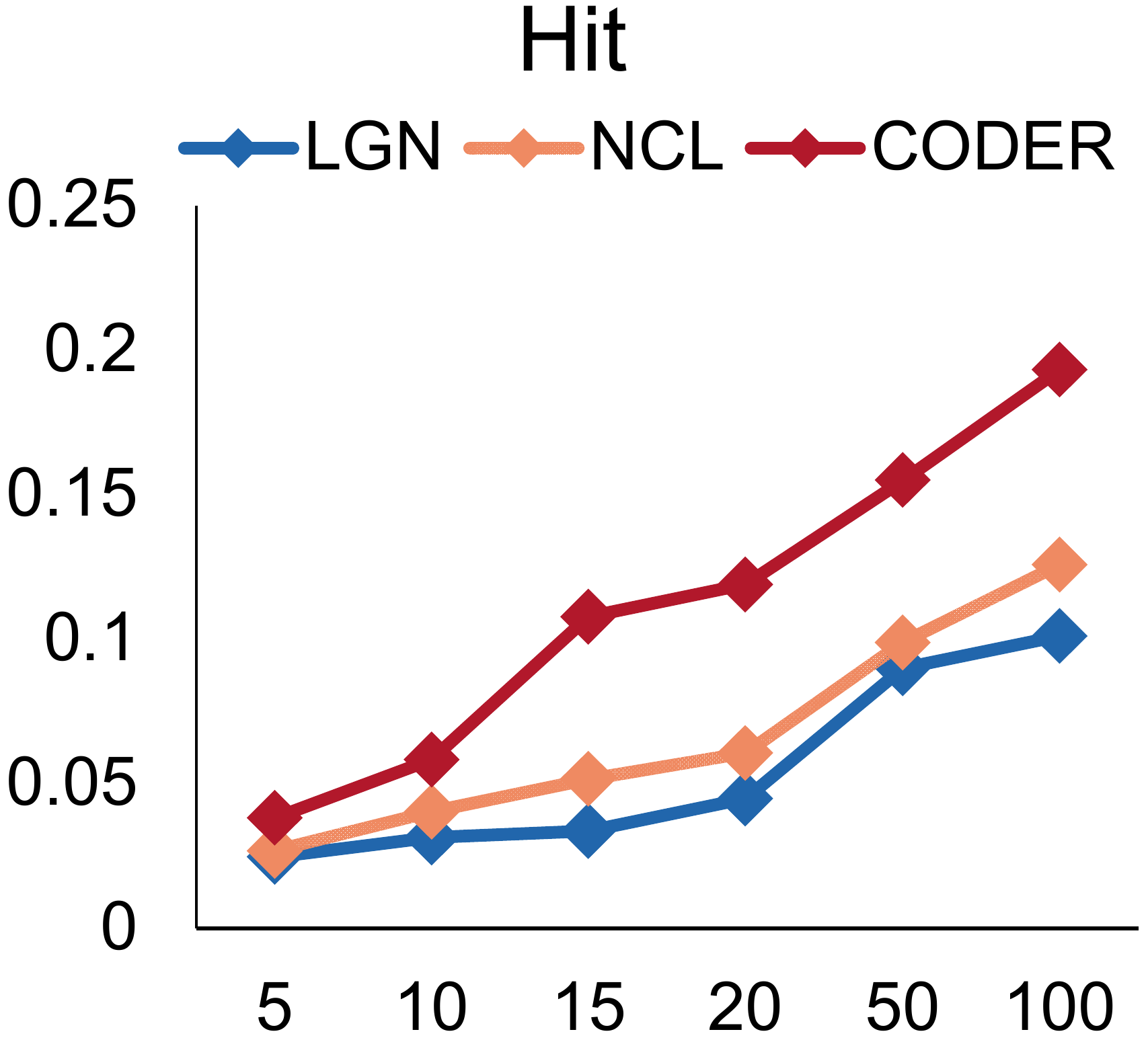}\quad
  \includegraphics[width=0.45\linewidth]{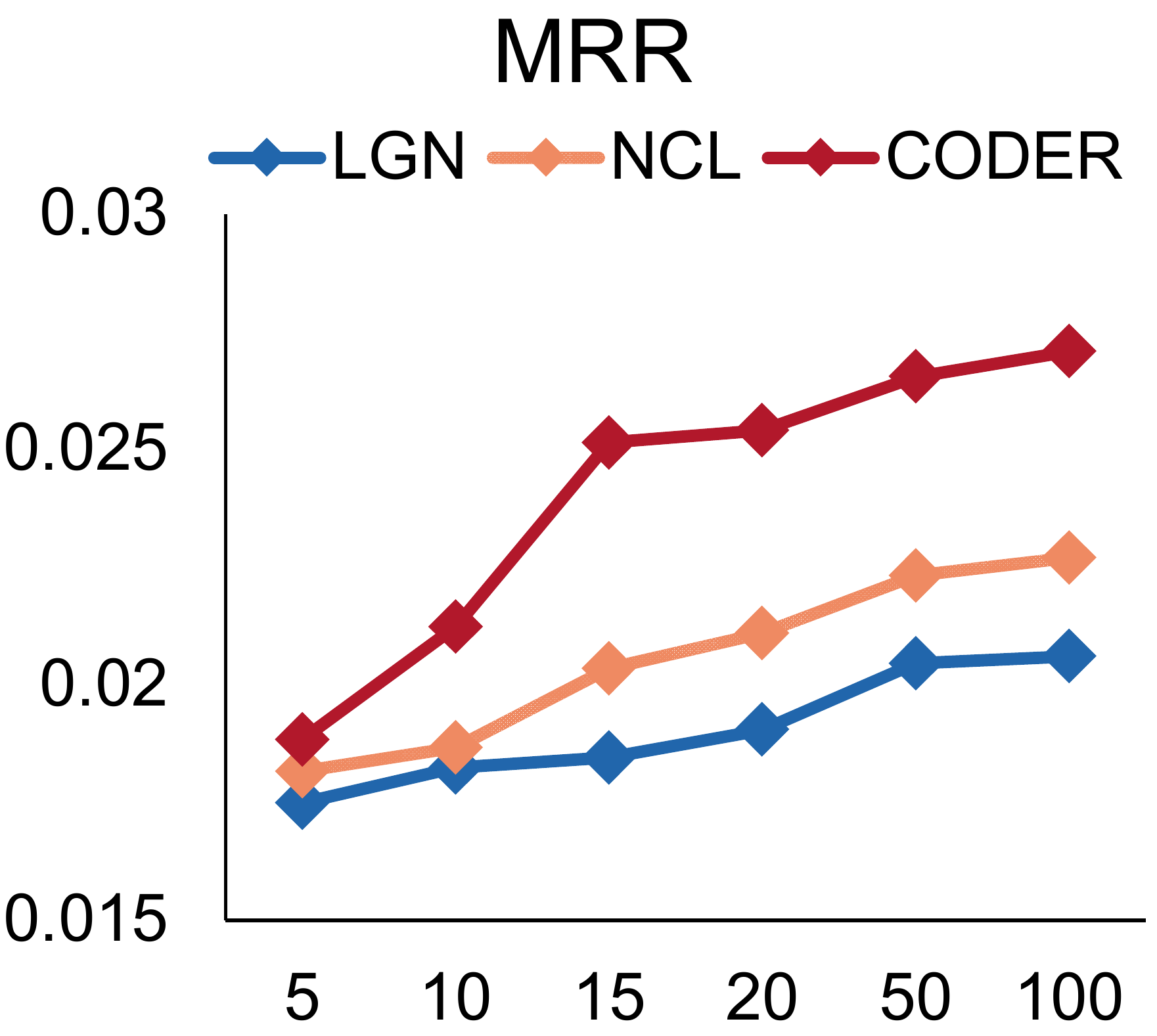}
  \caption{
  Cross-Project Performance of \model and the 2 strongest baselines under various $K, K \in [5, 100]$.
  }
  \label{fig:cross_project_rec}
\end{figure}

\subsubsection{Cold-Start Recommendation}
\label{sec:cold_start}
User contribution is usually sparse due to the considerable workload and voluntary nature of OSS development. In this sense, it is important to accurately capture the users' preferences with few observed interactions. Our model is thus designed according to this principle. 
We define cold-start users as users with $\le 2$ interactions in the training set. To evaluate the model performance with fewer interactions, we choose NDCG@$K$, Recall$@K$, and Hit@$K$, where $K \in \{ 3, 5\}$. The strongest 4 baselines in Tab.~\ref{tab:same_repo} are evaluated for comparison.

As observed from Tab.~\ref{tab:cold}, performance for cold-start users is worse than that for all users in Tab.~\ref{tab:same_repo}. 
Notably, \model is able to achieve even greater performance improvement over the baseline models. It can be attributed to the following aspects: 
1) \model learns more accurate representations by fusing the fine-grained semantics of project files with their interacted users, which facilitates the learning of user preferences even in the absence of dense user interactions. 
2) By explicitly modeling multiple types of project-level behaviors, \model effectively models the users' interests to complement the sparse file-level contribution relations, which is more effective than encoding the project-level interactions in the embedding space.

\subsubsection{Cross-Project Recommendation}
\label{sec:uninteracted}
Although 91\% developers in our dataset focused on 1 project throughout their development, active contributors can work on multiple projects. For these contributors, the project core team can recommend development tasks based on their development experiences in previous projects.

During evaluation, we rank the interactions in projects each user has not yet interacted with in the training set. 
This setting is considerably more challenging than intra-project recommendation since the candidate item pool is significantly larger. 
According to the results in Fig.~\ref{fig:cross_project_rec}, \model consistently achieves superior performance by a large margin with respect to the baselines, especially for $K \ge 20$. The results show that \model jointly learns inter-project differences to choose the correct repositories and characterize intra-project distinctions to recommend the correct files within the chosen repositories. 



\begin{table}
\setlength{\tabcolsep}{1.1mm}\caption{File-level link prediction results for cold-start users. ``LGN'' stands for the baseline ``LightGCN''. The best performance is marked in bold. The second best is underlined.}
\begin{tabular}{l|l|ccccccc|c} 
\hline
\hline
& Metric & NeuMF           & NGCF  & LGN   & NCL           & \model          & Impr.   \\ 
\hline
        & NDCG@3 & 0.059         & 0.067 & 0.068 & \underline{0.090} & \textbf{0.126} & 40.9\%  \\
        & Hit@3  & 0.106         & 0.123 & 0.161 & \underline{0.211} & \textbf{0.224} & 5.9\%   \\
\multirow{2}{*}{ML}     & MRR@3  & 0.081         & 0.087 & 0.089 & \underline{0.119} & \textbf{0.165} & 38.3\%  \\
        & NDCG@5 & 0.068         & 0.078 & 0.088 & \underline{0.105} & \textbf{0.132} & 25.8\%  \\
        & Hit@5  & 0.161         & 0.162 & 0.230 & \underline{0.261} & \textbf{0.273} & 4.8\%   \\
        & MRR@5  & 0.093         & 0.097 & 0.105 & \underline{0.130} & \textbf{0.177} & 36.0\%  \\ 
\hline
        & NDCG@3 & \underline{0.078} & 0.063 & 0.055 & 0.075         & \textbf{0.119} & 53.0\%  \\
        & Hit@3  & 0.152         & 0.114 & 0.128 & \underline{0.165} & \textbf{0.238} & 44.4\%  \\
\multirow{2}{*}{DB} & MRR@3  & \underline{0.102} & 0.089 & 0.070 & 0.095         & \textbf{0.157} & 54.0\%  \\
        & NDCG@5 & 0.086         & 0.061 & 0.064 & \underline{0.086} & \textbf{0.127} & 47.5\%  \\
        & Hit@5  & 0.195         & 0.132 & 0.165 & \underline{0.220} & \textbf{0.287} & 30.6\%  \\
        & MRR@5  & \underline{0.112} & 0.093 & 0.079 & 0.106         & \textbf{0.168} & 50.2\%  \\ 
\hline
        & NDCG@3 & 0.079         & 0.075 & 0.085 & \underline{0.092} & \textbf{0.128} & 38.7\%  \\
        & Hit@3  & 0.171         & 0.165 & 0.179 & \underline{0.179} & \textbf{0.242} & 35.6\%  \\
\multirow{2}{*}{FS} & MRR@3  & 0.110         & 0.095 & 0.104 & \underline{0.116} & \textbf{0.171} & 48.0\%  \\
        & NDCG@5 & 0.086         & 0.086 & 0.085 & \underline{0.095} & \textbf{0.137} & 44.3\%  \\
        & Hit@5  & \underline{0.230} & 0.220 & 0.202 & 0.222         & \textbf{0.313} & 36.2\%  \\
        & MRR@5  & 0.124         & 0.106 & 0.109 & \underline{0.125} & \textbf{0.187} & 49.4\%  \\
\hline
\hline
\end{tabular}
\label{tab:cold}
 \vspace{-6mm}
\end{table}

\subsubsection{Ablation Studies}

In Fig.~\ref{fig:ablation_ml}, we compare the performance of our model (abbreviated as CD) among its 5 variants.
\textbf{CD-F} removes the code-user modality fusion strategy in Eq.~(\ref{eq:coattention_affinity_matrix}). 
\textbf{CD-C} excludes the structural contrastive learning objective in Eqs.~(\ref{eq:structure_loss_user}, \ref{eq:structure_loss_file}).
\textbf{CD-E} does not use the pretrained CodeBERT embeddings and instead applies TF-IDF encoding on the source code, a common approach in project recommendation models~\cite{xu2017scalable}.
\textbf{CD-P} removes the project-level aggregation in 
Sec.~\ref{sec:inter_level_aggregation}. 
\textbf{CD-S} disables the structural-level aggregation in Sec.~\ref{sec:structure_level_aggregation}. 
Results on the ML dataset are in Fig.~\ref{fig:ablation_ml}.

We observe that all 6 variants of \model outperform NCL, among which the full model (CD) performs the best, indicating the importance of each component in our model design.
The performance drops most significantly when we disable project-level aggregation in \textbf{CD-P}, indicating the importance of explicitly modeling user-project interactions through graph structures. 
We also observe a considerable decrease when we remove the structural-level aggregation (\textbf{CD-S}), implying that the structural information of files has a significant contribution towards the file representation.
\textbf{CD-E} does not lead to a more significant performance decrease, but is outperformed by \textbf{CD-F} where fine-grained code representations are present. Thus, user behaviors and project structural clues are more important than semantic features in code recommendation.

\section{Related Work}

\subsection{Research in Open Source Code}

Open sourcing has grown into a standard practice for software engineering~\cite{jiang2017open} and attract researchers to study social coding~\cite{yu2016reviewer}. 
Analytical studies focus on users' motivation~\cite{ye2003toward, gerosa2021shifting}, expertise~\cite{venkataramani2013discovery}, collaboration patterns~\cite{nahar2022collaboration}, and factors that impact the popularity~\cite{borges2016understanding} of projects. 
Methodological studies explore project classification~\cite{zhang2019higitclass} and code search~\cite{luan2019aroma}, connecting publications with projects~\cite{shao2020paper2repo}. 
Recently, the field of Neural Code Intelligence (NCI), which uses deep learning techniques to tackle analytical tasks on source code, has emerged as a promising direction to improve programming efficiency and reducing human errors within the software industry~\cite{Xu:2022ug}.
Although previous works explored the recommendation task in OSS development settings such as automatic suggestions of API function calls~\cite{nguyen2019focus, he2021pyart}, Good First Issues~\cite{xiao2022recommending,alderliesten2021initial}, and data preparation steps~\cite{yan2020auto}, no previous works have studied the challenging task of code recommendation task, which requires in-depth understanding of diverse user-item interactions and OSS projects written in multiple programming languages. 

\subsection{Recommender Systems}
The advances in deep learning greatly facilitate the evolution of recommender systems~\cite{chen2022learning, yuan2020future, hao2020p, zhang2021mining, oh2022implicit, shalaby2022m2trec}. 
Motivated by the success of Graph Neural Networks (GNN)~\cite{zhao2021heterogeneous, zhao2021multi, wang2022augmentation, zhu2021deep, zhu2021graph, mao2021source, freitas2022graph}, a series of graph-based recommender systems~\cite{wu2019session, li2022hypercomplex, wang2022multi} are proposed, which organize user behaviors into heterogeneous interaction graphs. These methods formulate item recommendation as link prediction or representation learning tasks~\cite{wang2021learning, zheng2021dgcn, sharma2022survey}, and utilize high-order relationships to infer user preferences, item attributes, and collaborative filtering signals~\cite{ying2018graph, shao2020paper2repo, wu2019neural}. 
Noticeably, traditional recommendation models 
cannot be easily transferred to code recommendation as they do not model unique signals in OSS development, such as project hierarchies and code semantics.


\section{Conclusion and Future Works}

We are the first to formulate the task of code recommendation for open source developers. We propose \model, a code recommendation model for open source projects written in diverse languages. 
Currently, our approach only considers recommending existing files to users. As \model harnesses the metadata and semantic features of files, it cannot deal with users creating new files where such information of the candidate files is absent. We plan to generalize our framework by 
allowing users to initialize files under their interested subdirectories.  
Moreover, our source code encoding scheme can be further improved by harnessing knowledge about programming languages such as using Abstract Syntax Tree (AST)~\cite{nguyen2016api} and data flow~\cite{he2021pyart, guo2021graphcodebert} (graphs that represent dependency relation between variables). 
Our current encoding scheme is a computationally efficient way to deal with the diversity of programming languages. In the future, we plan to incorporate such domain knowledge to improve the file representations at a finer granularity.
Our current encoding scheme is a computationally efficient way to deal with the diversity of programming languages. 
Moreover, the user representations can be further enhanced by modeling users' social relations~\cite{fan2019graph, mei2022mutually} and behaviors~\cite{jin2022towards, yang2022reinforcement, xu2022mining}.

\begin{acks}
This work was partially supported by NSF 2211557, NSF 1937599, NSF 2119643, NASA, SRC, Okawa Foundation Grant, Amazon Research Awards, Cisco research grant, Picsart Gifts, and Snapchat Gifts.
\end{acks}

\bibliographystyle{ACM-Reference-Format}
\bibliography{sample-base}


\begin{thebibliography}{81}


\ifx \showCODEN    \undefined \def \showCODEN     #1{\unskip}     \fi
\ifx \showDOI      \undefined \def \showDOI       #1{#1}\fi
\ifx \showISBNx    \undefined \def \showISBNx     #1{\unskip}     \fi
\ifx \showISBNxiii \undefined \def \showISBNxiii  #1{\unskip}     \fi
\ifx \showISSN     \undefined \def \showISSN      #1{\unskip}     \fi
\ifx \showLCCN     \undefined \def \showLCCN      #1{\unskip}     \fi
\ifx \shownote     \undefined \def \shownote      #1{#1}          \fi
\ifx \showarticletitle \undefined \def \showarticletitle #1{#1}   \fi
\ifx \showURL      \undefined \def \showURL       {\relax}        \fi
\providecommand\bibfield[2]{#2}
\providecommand\bibinfo[2]{#2}
\providecommand\natexlab[1]{#1}
\providecommand\showeprint[2][]{arXiv:#2}

\bibitem[Alderliesten and Zaidman(2021)]%
        {alderliesten2021initial}
\bibfield{author}{\bibinfo{person}{Jan Willem~David Alderliesten} {and}
  \bibinfo{person}{Andy Zaidman}.} \bibinfo{year}{2021}\natexlab{}.
\newblock \showarticletitle{An Initial Exploration of the “Good First
  Issue” Label for Newcomer Developers}. In
  \bibinfo{booktitle}{\emph{CHASE}}. \bibinfo{pages}{117--118}.
\newblock


\bibitem[Borges et~al\mbox{.}(2016)]%
        {borges2016understanding}
\bibfield{author}{\bibinfo{person}{Hudson Borges}, \bibinfo{person}{Andre
  Hora}, {and} \bibinfo{person}{Marco~Tulio Valente}.}
  \bibinfo{year}{2016}\natexlab{}.
\newblock \showarticletitle{Understanding the factors that impact the
  popularity of GitHub repositories}. In \bibinfo{booktitle}{\emph{ICSME}}.
  \bibinfo{pages}{334--344}.
\newblock


\bibitem[Chen et~al\mbox{.}(2022)]%
        {chen2022learning}
\bibfield{author}{\bibinfo{person}{Jin Chen}, \bibinfo{person}{Defu Lian},
  \bibinfo{person}{Binbin Jin}, \bibinfo{person}{Kai Zheng}, {and}
  \bibinfo{person}{Enhong Chen}.} \bibinfo{year}{2022}\natexlab{}.
\newblock \showarticletitle{Learning Recommenders for Implicit Feedback with
  Importance Resampling}. In \bibinfo{booktitle}{\emph{WWW}}.
  \bibinfo{pages}{1997--2005}.
\newblock


\bibitem[Coelho et~al\mbox{.}(2020)]%
        {coelho2020github}
\bibfield{author}{\bibinfo{person}{Jailton Coelho},
  \bibinfo{person}{Marco~Tulio Valente}, \bibinfo{person}{Luciano Milen}, {and}
  \bibinfo{person}{Luciana~L Silva}.} \bibinfo{year}{2020}\natexlab{}.
\newblock \showarticletitle{Is this GitHub project maintained? Measuring the
  level of maintenance activity of open-source projects}.
\newblock \bibinfo{journal}{\emph{Information and Software Technology}}
  \bibinfo{volume}{122} (\bibinfo{year}{2020}), \bibinfo{pages}{106274}.
\newblock


\bibitem[Di~Cosmo and Zacchiroli(2017)]%
        {di2017software}
\bibfield{author}{\bibinfo{person}{Roberto Di~Cosmo} {and}
  \bibinfo{person}{Stefano Zacchiroli}.} \bibinfo{year}{2017}\natexlab{}.
\newblock \showarticletitle{Software Heritage: Why and How to Preserve Software
  Source Code}. In \bibinfo{booktitle}{\emph{iPRES}}. \bibinfo{pages}{1--10}.
\newblock


\bibitem[Fan et~al\mbox{.}(2019)]%
        {fan2019graph}
\bibfield{author}{\bibinfo{person}{Wenqi Fan}, \bibinfo{person}{Yao Ma},
  \bibinfo{person}{Qing Li}, \bibinfo{person}{Yuan He}, \bibinfo{person}{Eric
  Zhao}, \bibinfo{person}{Jiliang Tang}, {and} \bibinfo{person}{Dawei Yin}.}
  \bibinfo{year}{2019}\natexlab{}.
\newblock \showarticletitle{Graph neural networks for social recommendation}.
  In \bibinfo{booktitle}{\emph{WWW}}. \bibinfo{pages}{417--426}.
\newblock


\bibitem[Feng et~al\mbox{.}(2020)]%
        {feng2020codebert}
\bibfield{author}{\bibinfo{person}{Zhangyin Feng}, \bibinfo{person}{Daya Guo},
  \bibinfo{person}{Duyu Tang}, \bibinfo{person}{Nan Duan},
  \bibinfo{person}{Xiaocheng Feng}, \bibinfo{person}{Ming Gong},
  \bibinfo{person}{Linjun Shou}, \bibinfo{person}{Bing Qin},
  \bibinfo{person}{Ting Liu}, \bibinfo{person}{Daxin Jiang}, {et~al\mbox{.}}}
  \bibinfo{year}{2020}\natexlab{}.
\newblock \showarticletitle{CodeBERT: A Pre-Trained Model for Programming and
  Natural Languages}. In \bibinfo{booktitle}{\emph{Findings of EMNLP 2020}}.
  \bibinfo{pages}{1536--1547}.
\newblock


\bibitem[Fey and Lenssen(2019)]%
        {Fey/Lenssen/2019}
\bibfield{author}{\bibinfo{person}{Matthias Fey} {and} \bibinfo{person}{Jan~E.
  Lenssen}.} \bibinfo{year}{2019}\natexlab{}.
\newblock \showarticletitle{Fast Graph Representation Learning with {PyTorch
  Geometric}}. In \bibinfo{booktitle}{\emph{RLGM@ICLR}}.
\newblock


\bibitem[Freitas et~al\mbox{.}(2022)]%
        {freitas2022graph}
\bibfield{author}{\bibinfo{person}{Scott Freitas}, \bibinfo{person}{Diyi Yang},
  \bibinfo{person}{Srijan Kumar}, \bibinfo{person}{Hanghang Tong}, {and}
  \bibinfo{person}{Duen~Horng Chau}.} \bibinfo{year}{2022}\natexlab{}.
\newblock \showarticletitle{Graph vulnerability and robustness: A survey}.
\newblock \bibinfo{journal}{\emph{TKDE}} (\bibinfo{year}{2022}).
\newblock


\bibitem[Gerosa et~al\mbox{.}(2021)]%
        {gerosa2021shifting}
\bibfield{author}{\bibinfo{person}{Marco Gerosa}, \bibinfo{person}{Igor Wiese},
  \bibinfo{person}{Bianca Trinkenreich}, \bibinfo{person}{Georg Link},
  \bibinfo{person}{Gregorio Robles}, \bibinfo{person}{Christoph Treude},
  \bibinfo{person}{Igor Steinmacher}, {and} \bibinfo{person}{Anita Sarma}.}
  \bibinfo{year}{2021}\natexlab{}.
\newblock \showarticletitle{The shifting sands of motivation: Revisiting what
  drives contributors in open source}. In \bibinfo{booktitle}{\emph{ICSE}}.
  \bibinfo{pages}{1046--1058}.
\newblock


\bibitem[GitHub(2016)]%
        {octoverse2016}
\bibfield{author}{\bibinfo{person}{GitHub}.} \bibinfo{year}{2016}\natexlab{}.
\newblock \bibinfo{booktitle}{\emph{The State of the Octoverse}}.
\newblock
\urldef\tempurl%
\url{https://octoverse.github.com/2016/}
\showURL{%
\tempurl}


\bibitem[GitHub(2022a)]%
        {githublang2022}
\bibfield{author}{\bibinfo{person}{GitHub}.} \bibinfo{year}{2022}\natexlab{a}.
\newblock \bibinfo{booktitle}{\emph{Collection: Programming Languages}}.
\newblock
\urldef\tempurl%
\url{https://github.com/collections/programming-languages}
\showURL{%
\tempurl}


\bibitem[GitHub(2022b)]%
        {githubarchive}
\bibfield{author}{\bibinfo{person}{GitHub}.} \bibinfo{year}{2022}\natexlab{b}.
\newblock \bibinfo{title}{Github Number of Repositories}.
\newblock
\newblock
\newblock
\shownote{\url{https://github.com/search}}.


\bibitem[Glorot and Bengio(2010)]%
        {glorot2010understanding}
\bibfield{author}{\bibinfo{person}{Xavier Glorot} {and} \bibinfo{person}{Yoshua
  Bengio}.} \bibinfo{year}{2010}\natexlab{}.
\newblock \showarticletitle{Understanding the difficulty of training deep
  feedforward neural networks}. In \bibinfo{booktitle}{\emph{AISTATS}}.
  \bibinfo{pages}{249--256}.
\newblock


\bibitem[Guo et~al\mbox{.}(2021)]%
        {guo2021graphcodebert}
\bibfield{author}{\bibinfo{person}{Daya Guo}, \bibinfo{person}{Shuo Ren},
  \bibinfo{person}{Shuai Lu}, \bibinfo{person}{Zhangyin Feng},
  \bibinfo{person}{Duyu Tang}, \bibinfo{person}{Shujie Liu},
  \bibinfo{person}{Long Zhou}, \bibinfo{person}{Nan Duan},
  \bibinfo{person}{Alexey Svyatkovskiy}, \bibinfo{person}{Shengyu Fu},
  {et~al\mbox{.}}} \bibinfo{year}{2021}\natexlab{}.
\newblock \showarticletitle{GraphCodeBERT: Pre-training Code Representations
  with Data Flow}. In \bibinfo{booktitle}{\emph{ICLR}}.
\newblock


\bibitem[Hamilton et~al\mbox{.}(2017)]%
        {hamilton2017inductive}
\bibfield{author}{\bibinfo{person}{Will Hamilton}, \bibinfo{person}{Zhitao
  Ying}, {and} \bibinfo{person}{Jure Leskovec}.}
  \bibinfo{year}{2017}\natexlab{}.
\newblock \showarticletitle{Inductive representation learning on large graphs}.
\newblock \bibinfo{journal}{\emph{NIPS}}  \bibinfo{volume}{30}
  (\bibinfo{year}{2017}).
\newblock


\bibitem[Hao et~al\mbox{.}(2020)]%
        {hao2020p}
\bibfield{author}{\bibinfo{person}{Junheng Hao}, \bibinfo{person}{Tong Zhao},
  \bibinfo{person}{Jin Li}, \bibinfo{person}{Xin~Luna Dong},
  \bibinfo{person}{Christos Faloutsos}, \bibinfo{person}{Yizhou Sun}, {and}
  \bibinfo{person}{Wei Wang}.} \bibinfo{year}{2020}\natexlab{}.
\newblock \showarticletitle{P-companion: A principled framework for diversified
  complementary product recommendation}. In \bibinfo{booktitle}{\emph{CIKM}}.
  \bibinfo{pages}{2517--2524}.
\newblock


\bibitem[He et~al\mbox{.}(2020)]%
        {he2020lightgcn}
\bibfield{author}{\bibinfo{person}{Xiangnan He}, \bibinfo{person}{Kuan Deng},
  \bibinfo{person}{Xiang Wang}, \bibinfo{person}{Yan Li},
  \bibinfo{person}{Yongdong Zhang}, {and} \bibinfo{person}{Meng Wang}.}
  \bibinfo{year}{2020}\natexlab{}.
\newblock \showarticletitle{Lightgcn: Simplifying and powering graph
  convolution network for recommendation}. In
  \bibinfo{booktitle}{\emph{SIGIR}}. \bibinfo{pages}{639--648}.
\newblock


\bibitem[He et~al\mbox{.}(2017)]%
        {he2017neural}
\bibfield{author}{\bibinfo{person}{Xiangnan He}, \bibinfo{person}{Lizi Liao},
  \bibinfo{person}{Hanwang Zhang}, \bibinfo{person}{Liqiang Nie},
  \bibinfo{person}{Xia Hu}, {and} \bibinfo{person}{Tat-Seng Chua}.}
  \bibinfo{year}{2017}\natexlab{}.
\newblock \showarticletitle{Neural collaborative filtering}. In
  \bibinfo{booktitle}{\emph{WWW}}. \bibinfo{pages}{173--182}.
\newblock


\bibitem[He et~al\mbox{.}(2021)]%
        {he2021pyart}
\bibfield{author}{\bibinfo{person}{Xincheng He}, \bibinfo{person}{Lei Xu},
  \bibinfo{person}{Xiangyu Zhang}, \bibinfo{person}{Rui Hao},
  \bibinfo{person}{Yang Feng}, {and} \bibinfo{person}{Baowen Xu}.}
  \bibinfo{year}{2021}\natexlab{}.
\newblock \showarticletitle{Pyart: Python api recommendation in real-time}. In
  \bibinfo{booktitle}{\emph{ICSE}}. \bibinfo{pages}{1634--1645}.
\newblock


\bibitem[Hu et~al\mbox{.}(2018)]%
        {hu2018summarizing}
\bibfield{author}{\bibinfo{person}{Xing Hu}, \bibinfo{person}{Ge Li},
  \bibinfo{person}{Xin Xia}, \bibinfo{person}{David Lo}, \bibinfo{person}{Shuai
  Lu}, {and} \bibinfo{person}{Zhi Jin}.} \bibinfo{year}{2018}\natexlab{}.
\newblock \showarticletitle{Summarizing source code with transferred API
  knowledge}. In \bibinfo{booktitle}{\emph{IJCAI}}.
  \bibinfo{pages}{2269--2275}.
\newblock


\bibitem[Husain et~al\mbox{.}(2019)]%
        {husain2019codesearchnet}
\bibfield{author}{\bibinfo{person}{Hamel Husain}, \bibinfo{person}{Ho-Hsiang
  Wu}, \bibinfo{person}{Tiferet Gazit}, \bibinfo{person}{Miltiadis Allamanis},
  {and} \bibinfo{person}{Marc Brockschmidt}.} \bibinfo{year}{2019}\natexlab{}.
\newblock \showarticletitle{Codesearchnet challenge: Evaluating the state of
  semantic code search}.
\newblock \bibinfo{journal}{\emph{arXiv preprint arXiv:1909.09436}}
  (\bibinfo{year}{2019}).
\newblock


\bibitem[Jiang et~al\mbox{.}(2017)]%
        {jiang2017open}
\bibfield{author}{\bibinfo{person}{Jyun-Yu Jiang}, \bibinfo{person}{Pu-Jen
  Cheng}, {and} \bibinfo{person}{Wei Wang}.} \bibinfo{year}{2017}\natexlab{}.
\newblock \showarticletitle{Open source repository recommendation in social
  coding}. In \bibinfo{booktitle}{\emph{SIGIR}}. \bibinfo{pages}{1173--1176}.
\newblock


\bibitem[Jin et~al\mbox{.}(2023)]%
        {jin2022prototypical}
\bibfield{author}{\bibinfo{person}{Yiqiao Jin}, \bibinfo{person}{Xiting Wang},
  \bibinfo{person}{Yaru Hao}, \bibinfo{person}{Yizhou Sun}, {and}
  \bibinfo{person}{Xing Xie}.} \bibinfo{year}{2023}\natexlab{}.
\newblock \showarticletitle{Prototypical Fine-tuning: Towards Robust
  Performance Under Varying Data Sizes}. In \bibinfo{booktitle}{\emph{AAAI}}.
\newblock


\bibitem[Jin et~al\mbox{.}(2022)]%
        {jin2022towards}
\bibfield{author}{\bibinfo{person}{Yiqiao Jin}, \bibinfo{person}{Xiting Wang},
  \bibinfo{person}{Ruichao Yang}, \bibinfo{person}{Yizhou Sun},
  \bibinfo{person}{Wei Wang}, \bibinfo{person}{Hao Liao}, {and}
  \bibinfo{person}{Xing Xie}.} \bibinfo{year}{2022}\natexlab{}.
\newblock \showarticletitle{Towards fine-grained reasoning for fake news
  detection}. In \bibinfo{booktitle}{\emph{AAAI}}, Vol.~\bibinfo{volume}{36}.
  \bibinfo{pages}{5746--5754}.
\newblock


\bibitem[Kenton and Toutanova(2019)]%
        {kenton2019bert}
\bibfield{author}{\bibinfo{person}{Jacob Devlin Ming-Wei~Chang Kenton} {and}
  \bibinfo{person}{Lee~Kristina Toutanova}.} \bibinfo{year}{2019}\natexlab{}.
\newblock \showarticletitle{BERT: Pre-training of Deep Bidirectional
  Transformers for Language Understanding}. In
  \bibinfo{booktitle}{\emph{NAACL}}.
\newblock


\bibitem[Kingma and Ba(2015)]%
        {kingma2015adam}
\bibfield{author}{\bibinfo{person}{Diederik~P Kingma} {and}
  \bibinfo{person}{Jimmy Ba}.} \bibinfo{year}{2015}\natexlab{}.
\newblock \showarticletitle{Adam: A Method for Stochastic Optimization}. In
  \bibinfo{booktitle}{\emph{ICLR}}.
\newblock


\bibitem[Kipf and Welling(2016)]%
        {kipf2016semi}
\bibfield{author}{\bibinfo{person}{Thomas~N Kipf} {and} \bibinfo{person}{Max
  Welling}.} \bibinfo{year}{2016}\natexlab{}.
\newblock \showarticletitle{Semi-supervised classification with graph
  convolutional networks}. In \bibinfo{booktitle}{\emph{ICLR}}.
\newblock


\bibitem[Li et~al\mbox{.}(2022)]%
        {li2022hypercomplex}
\bibfield{author}{\bibinfo{person}{Anchen Li}, \bibinfo{person}{Bo Yang},
  \bibinfo{person}{Huan Huo}, {and} \bibinfo{person}{Farookh Hussain}.}
  \bibinfo{year}{2022}\natexlab{}.
\newblock \showarticletitle{Hypercomplex Graph Collaborative Filtering}. In
  \bibinfo{booktitle}{\emph{WWW}}. \bibinfo{pages}{1914--1922}.
\newblock


\bibitem[Lin et~al\mbox{.}(2022)]%
        {lin2022improving}
\bibfield{author}{\bibinfo{person}{Zihan Lin}, \bibinfo{person}{Changxin Tian},
  \bibinfo{person}{Yupeng Hou}, {and} \bibinfo{person}{Wayne~Xin Zhao}.}
  \bibinfo{year}{2022}\natexlab{}.
\newblock \showarticletitle{Improving Graph Collaborative Filtering with
  Neighborhood-enriched Contrastive Learning}. In
  \bibinfo{booktitle}{\emph{WWW}}. \bibinfo{pages}{2320--2329}.
\newblock


\bibitem[Liu et~al\mbox{.}(2019)]%
        {liu2019roberta}
\bibfield{author}{\bibinfo{person}{Yinhan Liu}, \bibinfo{person}{Myle Ott},
  \bibinfo{person}{Naman Goyal}, \bibinfo{person}{Jingfei Du},
  \bibinfo{person}{Mandar Joshi}, \bibinfo{person}{Danqi Chen},
  \bibinfo{person}{Omer Levy}, \bibinfo{person}{Mike Lewis},
  \bibinfo{person}{Luke Zettlemoyer}, {and} \bibinfo{person}{Veselin
  Stoyanov}.} \bibinfo{year}{2019}\natexlab{}.
\newblock \showarticletitle{Roberta: A robustly optimized bert pretraining
  approach}.
\newblock \bibinfo{journal}{\emph{arXiv preprint arXiv:1907.11692}}
  (\bibinfo{year}{2019}).
\newblock


\bibitem[Lu et~al\mbox{.}(2016)]%
        {lu2016hierarchical}
\bibfield{author}{\bibinfo{person}{Jiasen Lu}, \bibinfo{person}{Jianwei Yang},
  \bibinfo{person}{Dhruv Batra}, {and} \bibinfo{person}{Devi Parikh}.}
  \bibinfo{year}{2016}\natexlab{}.
\newblock \showarticletitle{Hierarchical question-image co-attention for visual
  question answering}.
\newblock \bibinfo{journal}{\emph{NIPS}}  \bibinfo{volume}{29}
  (\bibinfo{year}{2016}).
\newblock


\bibitem[Luan et~al\mbox{.}(2019)]%
        {luan2019aroma}
\bibfield{author}{\bibinfo{person}{Sifei Luan}, \bibinfo{person}{Di Yang},
  \bibinfo{person}{Celeste Barnaby}, \bibinfo{person}{Koushik Sen}, {and}
  \bibinfo{person}{Satish Chandra}.} \bibinfo{year}{2019}\natexlab{}.
\newblock \showarticletitle{Aroma: Code recommendation via structural code
  search}. In \bibinfo{booktitle}{\emph{OOPSLA}}.
\newblock


\bibitem[Mao et~al\mbox{.}(2021)]%
        {mao2021source}
\bibfield{author}{\bibinfo{person}{Haitao Mao}, \bibinfo{person}{Lun Du},
  \bibinfo{person}{Yujia Zheng}, \bibinfo{person}{Qiang Fu},
  \bibinfo{person}{Zelin Li}, \bibinfo{person}{Xu Chen}, \bibinfo{person}{Han
  Shi}, {and} \bibinfo{person}{Dongmei Zhang}.}
  \bibinfo{year}{2021}\natexlab{}.
\newblock \showarticletitle{Source free unsupervised graph domain adaptation}.
\newblock \bibinfo{journal}{\emph{arXiv preprint arXiv:2112.00955}}
  (\bibinfo{year}{2021}).
\newblock


\bibitem[McDonald and Goggins(2013)]%
        {mcdonald2013performance}
\bibfield{author}{\bibinfo{person}{Nora McDonald} {and} \bibinfo{person}{Sean
  Goggins}.} \bibinfo{year}{2013}\natexlab{}.
\newblock \showarticletitle{Performance and participation in open source
  software on github}.
\newblock In \bibinfo{booktitle}{\emph{CHI}}. \bibinfo{pages}{139--144}.
\newblock


\bibitem[Mei et~al\mbox{.}(2022)]%
        {mei2022mutually}
\bibfield{author}{\bibinfo{person}{Xin Mei}, \bibinfo{person}{Xiaoyan Cai},
  \bibinfo{person}{Sen Xu}, \bibinfo{person}{Wenjie Li},
  \bibinfo{person}{Shirui Pan}, {and} \bibinfo{person}{Libin Yang}.}
  \bibinfo{year}{2022}\natexlab{}.
\newblock \showarticletitle{Mutually reinforced network embedding: An
  integrated approach to research paper recommendation}.
\newblock \bibinfo{journal}{\emph{Expert Systems with Applications}}
  (\bibinfo{year}{2022}), \bibinfo{pages}{117616}.
\newblock


\bibitem[Miceli-Barone and Sennrich(2017)]%
        {miceli2017parallel}
\bibfield{author}{\bibinfo{person}{Antonio~Valerio Miceli-Barone} {and}
  \bibinfo{person}{Rico Sennrich}.} \bibinfo{year}{2017}\natexlab{}.
\newblock \showarticletitle{A Parallel Corpus of Python Functions and
  Documentation Strings for Automated Code Documentation and Code Generation}.
  In \bibinfo{booktitle}{\emph{IJCNLP}}. \bibinfo{pages}{314--319}.
\newblock


\bibitem[Nahar et~al\mbox{.}(2022)]%
        {nahar2022collaboration}
\bibfield{author}{\bibinfo{person}{Nadia Nahar}, \bibinfo{person}{Shurui Zhou},
  \bibinfo{person}{Grace Lewis}, {and} \bibinfo{person}{Christian
  K{\"a}stner}.} \bibinfo{year}{2022}\natexlab{}.
\newblock \showarticletitle{Collaboration Challenges in Building ML-Enabled
  Systems: Communication, Documentation, Engineering, and Process}.
\newblock \bibinfo{journal}{\emph{Organization}} \bibinfo{volume}{1},
  \bibinfo{number}{2} (\bibinfo{year}{2022}), \bibinfo{pages}{3}.
\newblock


\bibitem[Nguyen et~al\mbox{.}(2016)]%
        {nguyen2016api}
\bibfield{author}{\bibinfo{person}{Anh~Tuan Nguyen}, \bibinfo{person}{Michael
  Hilton}, \bibinfo{person}{Mihai Codoban}, \bibinfo{person}{Hoan~Anh Nguyen},
  \bibinfo{person}{Lily Mast}, \bibinfo{person}{Eli Rademacher},
  \bibinfo{person}{Tien~N Nguyen}, {and} \bibinfo{person}{Danny Dig}.}
  \bibinfo{year}{2016}\natexlab{}.
\newblock \showarticletitle{API code recommendation using statistical learning
  from fine-grained changes}. In \bibinfo{booktitle}{\emph{SIGSOFT}}.
  \bibinfo{pages}{511--522}.
\newblock


\bibitem[Nguyen et~al\mbox{.}(2019)]%
        {nguyen2019focus}
\bibfield{author}{\bibinfo{person}{Phuong~T Nguyen}, \bibinfo{person}{Juri
  Di~Rocco}, \bibinfo{person}{Davide Di~Ruscio}, \bibinfo{person}{Lina Ochoa},
  \bibinfo{person}{Thomas Degueule}, {and} \bibinfo{person}{Massimiliano
  Di~Penta}.} \bibinfo{year}{2019}\natexlab{}.
\newblock \showarticletitle{Focus: A recommender system for mining api function
  calls and usage patterns}. In \bibinfo{booktitle}{\emph{ICSE}}.
  \bibinfo{pages}{1050--1060}.
\newblock


\bibitem[Oh et~al\mbox{.}(2022)]%
        {oh2022implicit}
\bibfield{author}{\bibinfo{person}{Sejoon Oh}, \bibinfo{person}{Ankur
  Bhardwaj}, \bibinfo{person}{Jongseok Han}, \bibinfo{person}{Sungchul Kim},
  \bibinfo{person}{Ryan~A Rossi}, {and} \bibinfo{person}{Srijan Kumar}.}
  \bibinfo{year}{2022}\natexlab{}.
\newblock \showarticletitle{Implicit Session Contexts for Next-Item
  Recommendations}. In \bibinfo{booktitle}{\emph{CIKM}}.
  \bibinfo{pages}{4364--4368}.
\newblock


\bibitem[Paszke et~al\mbox{.}(2019)]%
        {paszke2019pytorch}
\bibfield{author}{\bibinfo{person}{Adam Paszke}, \bibinfo{person}{Sam Gross},
  \bibinfo{person}{Francisco Massa}, \bibinfo{person}{Adam Lerer},
  \bibinfo{person}{James Bradbury}, \bibinfo{person}{Gregory Chanan},
  \bibinfo{person}{Trevor Killeen}, \bibinfo{person}{Zeming Lin},
  \bibinfo{person}{Natalia Gimelshein}, \bibinfo{person}{Luca Antiga},
  {et~al\mbox{.}}} \bibinfo{year}{2019}\natexlab{}.
\newblock \showarticletitle{Pytorch: An imperative style, high-performance deep
  learning library}.
\newblock \bibinfo{journal}{\emph{NIPS}}  \bibinfo{volume}{32}.
\newblock


\bibitem[Rendle et~al\mbox{.}(2009)]%
        {rendle2009bpr}
\bibfield{author}{\bibinfo{person}{Steffen Rendle}, \bibinfo{person}{Christoph
  Freudenthaler}, \bibinfo{person}{Zeno Gantner}, {and} \bibinfo{person}{Lars
  Schmidt-Thieme}.} \bibinfo{year}{2009}\natexlab{}.
\newblock \showarticletitle{BPR: Bayesian personalized ranking from implicit
  feedback}. In \bibinfo{booktitle}{\emph{UAI}}. \bibinfo{pages}{452--461}.
\newblock


\bibitem[Shalaby et~al\mbox{.}(2022)]%
        {shalaby2022m2trec}
\bibfield{author}{\bibinfo{person}{Walid Shalaby}, \bibinfo{person}{Sejoon Oh},
  \bibinfo{person}{Amir Afsharinejad}, \bibinfo{person}{Srijan Kumar}, {and}
  \bibinfo{person}{Xiquan Cui}.} \bibinfo{year}{2022}\natexlab{}.
\newblock \showarticletitle{M2TRec: Metadata-aware Multi-task Transformer for
  Large-scale and Cold-start free Session-based Recommendations}. In
  \bibinfo{booktitle}{\emph{RecSys}}. \bibinfo{pages}{573--578}.
\newblock


\bibitem[Shao et~al\mbox{.}(2020)]%
        {shao2020paper2repo}
\bibfield{author}{\bibinfo{person}{Huajie Shao}, \bibinfo{person}{Dachun Sun},
  \bibinfo{person}{Jiahao Wu}, \bibinfo{person}{Zecheng Zhang},
  \bibinfo{person}{Aston Zhang}, \bibinfo{person}{Shuochao Yao},
  \bibinfo{person}{Shengzhong Liu}, \bibinfo{person}{Tianshi Wang},
  \bibinfo{person}{Chao Zhang}, {and} \bibinfo{person}{Tarek Abdelzaher}.}
  \bibinfo{year}{2020}\natexlab{}.
\newblock \showarticletitle{paper2repo: GitHub repository recommendation for
  academic papers}. In \bibinfo{booktitle}{\emph{WWW}}.
  \bibinfo{pages}{629--639}.
\newblock


\bibitem[Sharma et~al\mbox{.}(2022)]%
        {sharma2022survey}
\bibfield{author}{\bibinfo{person}{Kartik Sharma}, \bibinfo{person}{Yeon-Chang
  Lee}, \bibinfo{person}{Sivagami Nambi}, \bibinfo{person}{Aditya Salian},
  \bibinfo{person}{Shlok Shah}, \bibinfo{person}{Sang-Wook Kim}, {and}
  \bibinfo{person}{Srijan Kumar}.} \bibinfo{year}{2022}\natexlab{}.
\newblock \showarticletitle{A Survey of Graph Neural Networks for Social
  Recommender Systems}.
\newblock \bibinfo{journal}{\emph{arXiv preprint arXiv:2212.04481}}
  (\bibinfo{year}{2022}).
\newblock


\bibitem[Srivastava and Salakhutdinov(2012)]%
        {srivastava2012multimodal}
\bibfield{author}{\bibinfo{person}{Nitish Srivastava} {and}
  \bibinfo{person}{Russ~R Salakhutdinov}.} \bibinfo{year}{2012}\natexlab{}.
\newblock \showarticletitle{Multimodal learning with deep boltzmann machines}.
\newblock \bibinfo{journal}{\emph{NIPS}}.
\newblock


\bibitem[Steinmacher et~al\mbox{.}(2014)]%
        {steinmacher2014preliminary}
\bibfield{author}{\bibinfo{person}{Igor Steinmacher},
  \bibinfo{person}{Ana~Paula Chaves}, \bibinfo{person}{Tayana~Uchoa Conte},
  {and} \bibinfo{person}{Marco~Aur{\'e}lio Gerosa}.}
  \bibinfo{year}{2014}\natexlab{}.
\newblock \showarticletitle{Preliminary empirical identification of barriers
  faced by newcomers to Open Source Software projects}. In
  \bibinfo{booktitle}{\emph{SBES}}. \bibinfo{pages}{51--60}.
\newblock


\bibitem[Veli{\v{c}}kovi{\'c} et~al\mbox{.}(2018)]%
        {velivckovic2018graph}
\bibfield{author}{\bibinfo{person}{Petar Veli{\v{c}}kovi{\'c}},
  \bibinfo{person}{Guillem Cucurull}, \bibinfo{person}{Arantxa Casanova},
  \bibinfo{person}{Adriana Romero}, \bibinfo{person}{Pietro Li{\`o}}, {and}
  \bibinfo{person}{Yoshua Bengio}.} \bibinfo{year}{2018}\natexlab{}.
\newblock \showarticletitle{Graph Attention Networks}. In
  \bibinfo{booktitle}{\emph{ICLR}}.
\newblock


\bibitem[Venkataramani et~al\mbox{.}(2013)]%
        {venkataramani2013discovery}
\bibfield{author}{\bibinfo{person}{Rahul Venkataramani}, \bibinfo{person}{Atul
  Gupta}, \bibinfo{person}{Allahbaksh Asadullah}, \bibinfo{person}{Basavaraju
  Muddu}, {and} \bibinfo{person}{Vasudev Bhat}.}
  \bibinfo{year}{2013}\natexlab{}.
\newblock \showarticletitle{Discovery of technical expertise from open source
  code repositories}. In \bibinfo{booktitle}{\emph{WWW}}.
  \bibinfo{pages}{97--98}.
\newblock


\bibitem[Wan et~al\mbox{.}(2018)]%
        {wan2018improving}
\bibfield{author}{\bibinfo{person}{Yao Wan}, \bibinfo{person}{Zhou Zhao},
  \bibinfo{person}{Min Yang}, \bibinfo{person}{Guandong Xu},
  \bibinfo{person}{Haochao Ying}, \bibinfo{person}{Jian Wu}, {and}
  \bibinfo{person}{Philip~S Yu}.} \bibinfo{year}{2018}\natexlab{}.
\newblock \showarticletitle{Improving automatic source code summarization via
  deep reinforcement learning}. In \bibinfo{booktitle}{\emph{ASE}}.
  \bibinfo{pages}{397--407}.
\newblock


\bibitem[Wang et~al\mbox{.}(2022b)]%
        {wang2022augmentation}
\bibfield{author}{\bibinfo{person}{Haonan Wang}, \bibinfo{person}{Jieyu Zhang},
  \bibinfo{person}{Qi Zhu}, {and} \bibinfo{person}{Wei Huang}.}
  \bibinfo{year}{2022}\natexlab{b}.
\newblock \showarticletitle{Augmentation-free graph contrastive learning}.
\newblock \bibinfo{journal}{\emph{arXiv preprint arXiv:2204.04874}}
  (\bibinfo{year}{2022}).
\newblock


\bibitem[Wang et~al\mbox{.}(2019)]%
        {wang2019neural}
\bibfield{author}{\bibinfo{person}{Xiang Wang}, \bibinfo{person}{Xiangnan He},
  \bibinfo{person}{Meng Wang}, \bibinfo{person}{Fuli Feng}, {and}
  \bibinfo{person}{Tat-Seng Chua}.} \bibinfo{year}{2019}\natexlab{}.
\newblock \showarticletitle{Neural graph collaborative filtering}. In
  \bibinfo{booktitle}{\emph{SIGIR}}. \bibinfo{pages}{165--174}.
\newblock


\bibitem[Wang et~al\mbox{.}(2021)]%
        {wang2021learning}
\bibfield{author}{\bibinfo{person}{Xiang Wang}, \bibinfo{person}{Tinglin
  Huang}, \bibinfo{person}{Dingxian Wang}, \bibinfo{person}{Yancheng Yuan},
  \bibinfo{person}{Zhenguang Liu}, \bibinfo{person}{Xiangnan He}, {and}
  \bibinfo{person}{Tat-Seng Chua}.} \bibinfo{year}{2021}\natexlab{}.
\newblock \showarticletitle{Learning intents behind interactions with knowledge
  graph for recommendation}. In \bibinfo{booktitle}{\emph{WWW}}.
  \bibinfo{pages}{878--887}.
\newblock


\bibitem[Wang et~al\mbox{.}(2022a)]%
        {wang2022multi}
\bibfield{author}{\bibinfo{person}{Xiting Wang}, \bibinfo{person}{Kunpeng Liu},
  \bibinfo{person}{Dongjie Wang}, \bibinfo{person}{Le Wu},
  \bibinfo{person}{Yanjie Fu}, {and} \bibinfo{person}{Xing Xie}.}
  \bibinfo{year}{2022}\natexlab{a}.
\newblock \showarticletitle{Multi-level recommendation reasoning over knowledge
  graphs with reinforcement learning}. In \bibinfo{booktitle}{\emph{WWW}}.
  \bibinfo{pages}{2098--2108}.
\newblock


\bibitem[Wolf et~al\mbox{.}(2020)]%
        {wolf2020transformers}
\bibfield{author}{\bibinfo{person}{Thomas Wolf}, \bibinfo{person}{Lysandre
  Debut}, \bibinfo{person}{Victor Sanh}, \bibinfo{person}{Julien Chaumond},
  \bibinfo{person}{Clement Delangue}, \bibinfo{person}{Anthony Moi},
  \bibinfo{person}{Pierric Cistac}, \bibinfo{person}{Tim Rault},
  \bibinfo{person}{R{\'e}mi Louf}, \bibinfo{person}{Morgan Funtowicz},
  {et~al\mbox{.}}} \bibinfo{year}{2020}\natexlab{}.
\newblock \showarticletitle{Transformers: State-of-the-art natural language
  processing}. In \bibinfo{booktitle}{\emph{EMNLP}}. \bibinfo{pages}{38--45}.
\newblock


\bibitem[Wu et~al\mbox{.}(2022)]%
        {wu2022adversarial}
\bibfield{author}{\bibinfo{person}{Junfei Wu}, \bibinfo{person}{Weizhi Xu},
  \bibinfo{person}{Qiang Liu}, \bibinfo{person}{Shu Wu}, {and}
  \bibinfo{person}{Liang Wang}.} \bibinfo{year}{2022}\natexlab{}.
\newblock \showarticletitle{Adversarial Contrastive Learning for Evidence-aware
  Fake News Detection with Graph Neural Networks}.
\newblock \bibinfo{journal}{\emph{arXiv preprint arXiv:2210.05498}}
  (\bibinfo{year}{2022}).
\newblock


\bibitem[Wu et~al\mbox{.}(2019a)]%
        {wu2019neural}
\bibfield{author}{\bibinfo{person}{Le Wu}, \bibinfo{person}{Peijie Sun},
  \bibinfo{person}{Yanjie Fu}, \bibinfo{person}{Richang Hong},
  \bibinfo{person}{Xiting Wang}, {and} \bibinfo{person}{Meng Wang}.}
  \bibinfo{year}{2019}\natexlab{a}.
\newblock \showarticletitle{A neural influence diffusion model for social
  recommendation}. In \bibinfo{booktitle}{\emph{SIGIR}}.
  \bibinfo{pages}{235--244}.
\newblock


\bibitem[Wu et~al\mbox{.}(2019b)]%
        {wu2019session}
\bibfield{author}{\bibinfo{person}{Shu Wu}, \bibinfo{person}{Yuyuan Tang},
  \bibinfo{person}{Yanqiao Zhu}, \bibinfo{person}{Liang Wang},
  \bibinfo{person}{Xing Xie}, {and} \bibinfo{person}{Tieniu Tan}.}
  \bibinfo{year}{2019}\natexlab{b}.
\newblock \showarticletitle{Session-based recommendation with graph neural
  networks}. In \bibinfo{booktitle}{\emph{AAAI}}, Vol.~\bibinfo{volume}{33}.
  \bibinfo{pages}{346--353}.
\newblock


\bibitem[Xiao et~al\mbox{.}(2022)]%
        {xiao2022recommending}
\bibfield{author}{\bibinfo{person}{Wenxin Xiao}, \bibinfo{person}{Hao He},
  \bibinfo{person}{Weiwei Xu}, \bibinfo{person}{Xin Tan},
  \bibinfo{person}{Jinhao Dong}, {and} \bibinfo{person}{Minghui Zhou}.}
  \bibinfo{year}{2022}\natexlab{}.
\newblock \showarticletitle{Recommending good first issues in GitHub OSS
  projects}. In \bibinfo{booktitle}{\emph{ICSE}}. \bibinfo{pages}{1830--1842}.
\newblock


\bibitem[Xu et~al\mbox{.}(2018)]%
        {xu2018powerful}
\bibfield{author}{\bibinfo{person}{Keyulu Xu}, \bibinfo{person}{Weihua Hu},
  \bibinfo{person}{Jure Leskovec}, {and} \bibinfo{person}{Stefanie Jegelka}.}
  \bibinfo{year}{2018}\natexlab{}.
\newblock \showarticletitle{How Powerful are Graph Neural Networks?}. In
  \bibinfo{booktitle}{\emph{ICLR}}.
\newblock


\bibitem[Xu et~al\mbox{.}(2017)]%
        {xu2017scalable}
\bibfield{author}{\bibinfo{person}{Wenyuan Xu}, \bibinfo{person}{Xiaobing Sun},
  \bibinfo{person}{Xin Xia}, {and} \bibinfo{person}{Xiang Chen}.}
  \bibinfo{year}{2017}\natexlab{}.
\newblock \showarticletitle{Scalable relevant project recommendation on
  GitHub}. In \bibinfo{booktitle}{\emph{Internetware}}. \bibinfo{pages}{1--10}.
\newblock


\bibitem[Xu et~al\mbox{.}(2022)]%
        {xu2022mining}
\bibfield{author}{\bibinfo{person}{Weizhi Xu}, \bibinfo{person}{Junfei Wu},
  \bibinfo{person}{Qiang Liu}, \bibinfo{person}{Shu Wu}, {and}
  \bibinfo{person}{Liang Wang}.} \bibinfo{year}{2022}\natexlab{}.
\newblock \showarticletitle{Mining Fine-grained Semantics via Graph Neural
  Networks for Evidence-based Fake News Detection}.
\newblock \bibinfo{journal}{\emph{arXiv preprint arXiv:2201.06885}}
  (\bibinfo{year}{2022}).
\newblock


\bibitem[Xu and Zhu(2022)]%
        {Xu:2022ug}
\bibfield{author}{\bibinfo{person}{Yichen Xu} {and} \bibinfo{person}{Yanqiao
  Zhu}.} \bibinfo{year}{2022}\natexlab{}.
\newblock \showarticletitle{{A Survey on Pretrained Language Models for Neural
  Code Intelligence}}.
\newblock \bibinfo{journal}{\emph{arXiv.org}} (\bibinfo{date}{Dec.}
  \bibinfo{year}{2022}).
\newblock
\showeprint[arxiv]{2212.10079v1}~[cs.SE]


\bibitem[Yan and He(2020)]%
        {yan2020auto}
\bibfield{author}{\bibinfo{person}{Cong Yan} {and} \bibinfo{person}{Yeye He}.}
  \bibinfo{year}{2020}\natexlab{}.
\newblock \showarticletitle{Auto-suggest: Learning-to-recommend data
  preparation steps using data science notebooks}. In
  \bibinfo{booktitle}{\emph{SIGMOD}}. \bibinfo{pages}{1539--1554}.
\newblock


\bibitem[Yang et~al\mbox{.}(2022)]%
        {yang2022reinforcement}
\bibfield{author}{\bibinfo{person}{Ruichao Yang}, \bibinfo{person}{Xiting
  Wang}, \bibinfo{person}{Yiqiao Jin}, \bibinfo{person}{Chaozhuo Li},
  \bibinfo{person}{Jianxun Lian}, {and} \bibinfo{person}{Xing Xie}.}
  \bibinfo{year}{2022}\natexlab{}.
\newblock \showarticletitle{Reinforcement Subgraph Reasoning for Fake News
  Detection}. In \bibinfo{booktitle}{\emph{KDD}}. \bibinfo{pages}{2253--2262}.
\newblock


\bibitem[Ye and Kishida(2003)]%
        {ye2003toward}
\bibfield{author}{\bibinfo{person}{Yunwen Ye} {and} \bibinfo{person}{Kouichi
  Kishida}.} \bibinfo{year}{2003}\natexlab{}.
\newblock \showarticletitle{Toward an understanding of the motivation of open
  source software developers}. In \bibinfo{booktitle}{\emph{ICSE}}.
  \bibinfo{pages}{419--429}.
\newblock


\bibitem[Ying et~al\mbox{.}(2018)]%
        {ying2018graph}
\bibfield{author}{\bibinfo{person}{Rex Ying}, \bibinfo{person}{Ruining He},
  \bibinfo{person}{Kaifeng Chen}, \bibinfo{person}{Pong Eksombatchai},
  \bibinfo{person}{William~L Hamilton}, {and} \bibinfo{person}{Jure Leskovec}.}
  \bibinfo{year}{2018}\natexlab{}.
\newblock \showarticletitle{Graph convolutional neural networks for web-scale
  recommender systems}. In \bibinfo{booktitle}{\emph{KDD}}.
  \bibinfo{pages}{974--983}.
\newblock


\bibitem[Yu et~al\mbox{.}(2021)]%
        {yu2021graph}
\bibfield{author}{\bibinfo{person}{Xueli Yu}, \bibinfo{person}{Weizhi Xu},
  \bibinfo{person}{Zeyu Cui}, \bibinfo{person}{Shu Wu}, {and}
  \bibinfo{person}{Liang Wang}.} \bibinfo{year}{2021}\natexlab{}.
\newblock \showarticletitle{Graph-based Hierarchical Relevance Matching Signals
  for Ad-hoc Retrieval}. In \bibinfo{booktitle}{\emph{WWW}}.
  \bibinfo{pages}{778--787}.
\newblock


\bibitem[Yu et~al\mbox{.}(2016)]%
        {yu2016reviewer}
\bibfield{author}{\bibinfo{person}{Yue Yu}, \bibinfo{person}{Huaimin Wang},
  \bibinfo{person}{Gang Yin}, {and} \bibinfo{person}{Tao Wang}.}
  \bibinfo{year}{2016}\natexlab{}.
\newblock \showarticletitle{Reviewer recommendation for pull-requests in
  GitHub: What can we learn from code review and bug assignment?}
\newblock \bibinfo{journal}{\emph{Information and Software Technology}}
  \bibinfo{volume}{74} (\bibinfo{year}{2016}), \bibinfo{pages}{204--218}.
\newblock


\bibitem[Yu et~al\mbox{.}(2019)]%
        {yu2019deep}
\bibfield{author}{\bibinfo{person}{Zhou Yu}, \bibinfo{person}{Jun Yu},
  \bibinfo{person}{Yuhao Cui}, \bibinfo{person}{Dacheng Tao}, {and}
  \bibinfo{person}{Qi Tian}.} \bibinfo{year}{2019}\natexlab{}.
\newblock \showarticletitle{Deep modular co-attention networks for visual
  question answering}. In \bibinfo{booktitle}{\emph{CVPR}}.
  \bibinfo{pages}{6281--6290}.
\newblock


\bibitem[Yuan et~al\mbox{.}(2020)]%
        {yuan2020future}
\bibfield{author}{\bibinfo{person}{Fajie Yuan}, \bibinfo{person}{Xiangnan He},
  \bibinfo{person}{Haochuan Jiang}, \bibinfo{person}{Guibing Guo},
  \bibinfo{person}{Jian Xiong}, \bibinfo{person}{Zhezhao Xu}, {and}
  \bibinfo{person}{Yilin Xiong}.} \bibinfo{year}{2020}\natexlab{}.
\newblock \showarticletitle{Future data helps training: Modeling future
  contexts for session-based recommendation}. In
  \bibinfo{booktitle}{\emph{WWW}}. \bibinfo{pages}{303--313}.
\newblock


\bibitem[Zhang et~al\mbox{.}(2021)]%
        {zhang2021mining}
\bibfield{author}{\bibinfo{person}{Jinghao Zhang}, \bibinfo{person}{Yanqiao
  Zhu}, \bibinfo{person}{Qiang Liu}, \bibinfo{person}{Shu Wu},
  \bibinfo{person}{Shuhui Wang}, {and} \bibinfo{person}{Liang Wang}.}
  \bibinfo{year}{2021}\natexlab{}.
\newblock \showarticletitle{Mining Latent Structures for Multimedia
  Recommendation}. In \bibinfo{booktitle}{\emph{ACM MM}}.
  \bibinfo{pages}{3872--3880}.
\newblock


\bibitem[Zhang et~al\mbox{.}(2019)]%
        {zhang2019higitclass}
\bibfield{author}{\bibinfo{person}{Yu Zhang}, \bibinfo{person}{Frank~F Xu},
  \bibinfo{person}{Sha Li}, \bibinfo{person}{Yu Meng}, \bibinfo{person}{Xuan
  Wang}, \bibinfo{person}{Qi Li}, {and} \bibinfo{person}{Jiawei Han}.}
  \bibinfo{year}{2019}\natexlab{}.
\newblock \showarticletitle{Higitclass: Keyword-driven hierarchical
  classification of github repositories}. In \bibinfo{booktitle}{\emph{ICDM}}.
  \bibinfo{pages}{876--885}.
\newblock


\bibitem[Zhao et~al\mbox{.}(2021a)]%
        {zhao2021heterogeneous}
\bibfield{author}{\bibinfo{person}{Jianan Zhao}, \bibinfo{person}{Xiao Wang},
  \bibinfo{person}{Chuan Shi}, \bibinfo{person}{Binbin Hu},
  \bibinfo{person}{Guojie Song}, {and} \bibinfo{person}{Yanfang Ye}.}
  \bibinfo{year}{2021}\natexlab{a}.
\newblock \showarticletitle{Heterogeneous graph structure learning for graph
  neural networks}. In \bibinfo{booktitle}{\emph{AAAI}},
  Vol.~\bibinfo{volume}{35}. \bibinfo{pages}{4697--4705}.
\newblock


\bibitem[Zhao et~al\mbox{.}(2021b)]%
        {zhao2021multi}
\bibfield{author}{\bibinfo{person}{Jianan Zhao}, \bibinfo{person}{Qianlong
  Wen}, \bibinfo{person}{Shiyu Sun}, \bibinfo{person}{Yanfang Ye}, {and}
  \bibinfo{person}{Chuxu Zhang}.} \bibinfo{year}{2021}\natexlab{b}.
\newblock \showarticletitle{Multi-view Self-supervised Heterogeneous Graph
  Embedding}. In \bibinfo{booktitle}{\emph{ECML-PKDD}}.
  \bibinfo{pages}{319--334}.
\newblock


\bibitem[Zheng et~al\mbox{.}(2021a)]%
        {zheng2021dgcn}
\bibfield{author}{\bibinfo{person}{Yu Zheng}, \bibinfo{person}{Chen Gao},
  \bibinfo{person}{Liang Chen}, \bibinfo{person}{Depeng Jin}, {and}
  \bibinfo{person}{Yong Li}.} \bibinfo{year}{2021}\natexlab{a}.
\newblock \showarticletitle{DGCN: Diversified Recommendation with Graph
  Convolutional Networks}. In \bibinfo{booktitle}{\emph{WWW}}.
  \bibinfo{pages}{401--412}.
\newblock


\bibitem[Zheng et~al\mbox{.}(2021b)]%
        {zheng2021disentangling}
\bibfield{author}{\bibinfo{person}{Yu Zheng}, \bibinfo{person}{Chen Gao},
  \bibinfo{person}{Xiang Li}, \bibinfo{person}{Xiangnan He},
  \bibinfo{person}{Yong Li}, {and} \bibinfo{person}{Depeng Jin}.}
  \bibinfo{year}{2021}\natexlab{b}.
\newblock \showarticletitle{Disentangling user interest and conformity for
  recommendation with causal embedding}. In \bibinfo{booktitle}{\emph{WWW}}.
  \bibinfo{pages}{2980--2991}.
\newblock


\bibitem[Zhu et~al\mbox{.}(2014)]%
        {zhu2014patterns}
\bibfield{author}{\bibinfo{person}{Jiaxin Zhu}, \bibinfo{person}{Minghui Zhou},
  {and} \bibinfo{person}{Audris Mockus}.} \bibinfo{year}{2014}\natexlab{}.
\newblock \showarticletitle{Patterns of folder use and project popularity: A
  case study of GitHub repositories}. In \bibinfo{booktitle}{\emph{ESEM}}.
  \bibinfo{pages}{1--4}.
\newblock


\bibitem[Zhu et~al\mbox{.}(2021b)]%
        {zhu2021deep}
\bibfield{author}{\bibinfo{person}{Yanqiao Zhu}, \bibinfo{person}{Weizhi Xu},
  \bibinfo{person}{Jinghao Zhang}, \bibinfo{person}{Qiang Liu},
  \bibinfo{person}{Shu Wu}, {and} \bibinfo{person}{Liang Wang}.}
  \bibinfo{year}{2021}\natexlab{b}.
\newblock \showarticletitle{Deep graph structure learning for robust
  representations: A survey}.
\newblock \bibinfo{journal}{\emph{arXiv preprint arXiv:2103.03036}}
  (\bibinfo{year}{2021}).
\newblock


\bibitem[Zhu et~al\mbox{.}(2021a)]%
        {zhu2021graph}
\bibfield{author}{\bibinfo{person}{Yanqiao Zhu}, \bibinfo{person}{Yichen Xu},
  \bibinfo{person}{Feng Yu}, \bibinfo{person}{Qiang Liu}, \bibinfo{person}{Shu
  Wu}, {and} \bibinfo{person}{Liang Wang}.} \bibinfo{year}{2021}\natexlab{a}.
\newblock \showarticletitle{Graph contrastive learning with adaptive
  augmentation}. In \bibinfo{booktitle}{\emph{WWW}}.
  \bibinfo{pages}{2069--2080}.
\newblock


\end{thebibliography}


\appendix

\section{Datasets}
\label{supp:datasets}
We categorized the open source projects in our dataset using their GitHub topics. The topics used in each dataset are shown in Tab.~\ref{tab:github_topics}.

\begin{table}[h]
\caption{GitHub topics used during the construction of each dataset.}
\label{tab:github_topics}
\begin{tabular}{l|c}
\hline
\hline
Dataset     & Topics                                                                               \\ \hline
& \texttt{computer-vision}, \texttt{data-science}, \texttt{0deep-learning}, \\
ML  & \texttt{machine-learning}, \texttt{neural-network},  \\
 & \texttt{pytorch}, \texttt{tensorflow} \\ \hline
\multirow{2}{*}{FS} & \texttt{angular}, \texttt{css}, \texttt{html}, \texttt{javascript}, \texttt{js}, \\
& \texttt{nodejs}, \texttt{react}, \texttt{reactjs}, \texttt{vue}, \texttt{vuejs}                     \\ \hline
DB  & \texttt{database}, \texttt{graphql}, \texttt{mongodb}, \texttt{mysql}, \texttt{sql}                                             \\ \hline \hline
\end{tabular}
\end{table}

\section{Implementation Details}
\label{supp:implementation}

To derive $\mathbf{C}$ in Sec.~\ref{sec:code_user_modality_fusion}, we first convert the source code into a set of tokens, then partition them into code segments with equal number of tokens. As the source code is first tokenized and then partitioned at the token level, our partitioning scheme will not partition meaningful tokens in the middle. For example, if ``\texttt{def}'' is a meaningful token, our method will make sure that the token will not be split into multiple parts such as ``\texttt{de}'' and ``\texttt{f}''.

Next, we encode the tokens using \textsf{CodeBERT}. Each token produces a token-level representation. For each code segment $c_i$, we perform max pooling on all of its token representations to derive $\mathbf{c}_i$. 
We use \textsf{CodeBERT} to encode files written in all programming languages (PL) as well as natural language (NL), as \textsf{CodeBERT} is able to generalize to a wide range of programming languages and shows outstanding performances on NL-PL tasks.
As \textsf{CodeBERT} learns general-purpose code representations for both natural language (NL) and programming languages (PL), \textsf{CodeBERT} is suitable for encoding the rich natural language clues in OSS projects including the \texttt{README} file, which contains frequent context-switching between NL and PL.

\end{document}